\documentclass[peerreview]{IEEEtran}
\usepackage{cite}
\usepackage{graphicx}

\usepackage{algorithm}
\usepackage{algpseudocode}
\usepackage{amsmath}
\usepackage{amsfonts}
\usepackage{amssymb}

\usepackage{hyperref}
\usepackage[margin=1in]{geometry} 
\usepackage{booktabs} 
\usepackage{tabularx} 
\usepackage{listings}
\usepackage{longtable} 
\usepackage{multirow}
\usepackage{adjustbox}

\begin{document}
\title{Learning Automata-Based Enhancements to RPL: Pioneering Load-Balancing and Traffic Management in IoT}
\author{Mohammadhossein Homaei \\
\textit{Departamento de Ingeniería de Sistemas Informáticos y Telemáticos,} \\
\textit{Universidad de Extremadura, Av/ Universidad S/N, 10003, Cáceres, Extremadura, Spain} \\
\href{mailto:Homaei@ieee.org}{Email: Homaei@ieee.org}
}
\date{14 August 2024}

\maketitle

\begin{abstract}
The Internet of Things (IoT) signifies a revolutionary technological advancement, enhancing various applications through device interconnectivity while introducing significant challenges due to these devices' limited hardware and communication capabilities. To navigate these complexities, the Internet Engineering Task Force (IETF) has tailored the Routing Protocol for Low-Power and Lossy Networks (RPL) to meet the unique demands of IoT environments. However, RPL struggles with traffic congestion and load distribution issues, negatively impacting network performance and reliability. This paper presents a novel enhancement to RPL by integrating learning automata designed to optimize network traffic distribution. This enhanced protocol, the Learning Automata-based Load-Aware RPL (LALARPL), dynamically adjusts routing decisions based on real-time network conditions, achieving more effective load balancing and significantly reducing network congestion. Extensive simulations reveal that this approach outperforms existing methodologies, leading to notable improvements in packet delivery rates, end-to-end delay, and energy efficiency. The findings highlight the potential of our approach to enhance IoT network operations and extend the lifespan of network components. The effectiveness of learning automata in refining routing processes within RPL offers valuable insights that may drive future advancements in IoT networking, aiming for more robust, efficient, and sustainable network architectures.
\end{abstract}
\vspace{5mm} 

\section{Introduction}\label{sec1}
The Internet of Things (IoT) heralds an era of unprecedented connectivity, where the digital and physical realms converge through a vast network of embedded devices \cite{lakshmi2024computational_11, zainaddin2024recent_15, Homaei2024}. This paradigm encompasses a diverse range of applications, from enhancing the livability of smart cities and optimizing industrial processes to advancing healthcare outcomes and creating more efficient energy management systems. The proliferation of IoT devices, projected to exceed 75.44 billion by 2025, underscores the transformative potential of this technology to reshape our interactions with the world around us. Yet, as the fabric of the IoT continues to expand, it brings to the fore complex challenges that necessitate robust, scalable, and efficient networking protocols.

Low-Power and Lossy Networks (LLNs) are at the heart of IoT infrastructures, characterized by resource-constrained devices and inherently unreliable communication links \cite{darabkh2022rpl_07, Muzammal2021_55}. These networks are the backbone of IoT, facilitating the seamless data exchange among many connected devices. To cater to the unique demands of LLNs, the IETF introduced the Routing Protocol for Low-Power and Lossy Networks (RPL) \cite{rfc6553, almutairi2024survey_08, Estepa2021_60, Muzammal2021_64}. As the de facto standard for IoT networking, RPL is engineered to navigate the intricacies of LLNs, offering a flexible framework to support diverse traffic flows, including multipoint-to-point (MP2P), point-to-point (P2P), and point-to-multipoint (P2MP) communications. Despite its comprehensive design, RPL encounters formidable challenges that could stymie the IoT's growth and operational efficiency \cite{zormati2024review_09, Mishra2022_59}.

Among the myriad challenges RPL faces, congestion emerges as a critical bottleneck, severely impacting network performance \cite{tadigotla2020comprehensive_02, lamaazi2020comprehensive_13}. Congestion in IoT networks, particularly in RPL-based LLNs, manifests as accumulating data packets at certain nodes, leading to increased packet loss, latency, and energy consumption. This congestion is primarily attributed to IoT devices' high density and sporadic yet intense data transmission patterns, which overwhelm network resources. Furthermore, the inherent limitations of LLNs, such as restricted bandwidth and variable link quality, exacerbate the situation, making congestion control and load balancing paramount to the sustainability and reliability of IoT networks.

Effective congestion management and load balancing within RPL are not merely technical requisites but are imperative to realizing the full potential of the IoT \cite{maheshwari2022enhanced_14, Homaei2020_54, poorana2023protocol_37, Tarif2024, Tarif2024Enhance}. These mechanisms ensure equitable network traffic distribution, preventing overload situations and optimizing the use of scarce resources. By addressing congestion proactively, RPL can enhance the quality of service (QoS), extend network lifetime, and ensure reliable data delivery—factors crucial for the success of IoT applications, from critical healthcare monitoring systems to real-time industrial automation processes.

This paper explores enhancements to the LLNs, specifically within the RPL-class structure, by introducing the Learning Automata-based Load-Aware RPL (LALARPL) algorithm aimed at improving load balancing in IoT networks. Initially, the document highlights the primary challenges encountered in RPL, mainly focusing on congestion control across both broader IoT environments and specifically within RPL networks. Following this, it delves into the vital role and existing methods of load balancing in these networks (Section ~\ref{sec2}). The subsequent sections detail the Learning Automata framework, which is utilized to optimize routing decisions (Section ~\ref{sec3}), and provide a thorough description of the proposed LALARPL algorithm (Section ~\ref{sec4}). The effectiveness of LALARPL is rigorously tested through simulations that assess various performance metrics, including packet delivery ratio, throughput, fairness in throughput, end-to-end delay, energy consumption fairness, and network lifetime (Section ~\ref{sec5}). The paper concludes by synthesizing the findings, affirming the advantages of LALARPL in boosting network performance and reliability, and suggesting directions for future research. This includes the potential integration of mobility features to accommodate better dynamic network environments (Section ~\ref{sec6}). This structure ensures a systematic exploration of LALARPL's potential to revolutionize RPL implementations in IoT settings.

\begin{table}[H]
\caption{List of Abbreviations and their Definitions}
\label{tab:abbreviations}
\begin{scriptsize}
    
\centering
\begin{tabular}{|p{0.3\textwidth}|p{0.6\textwidth}|}
\hline
\textbf{Abbreviation} & \textbf{Definition} \\ \hline
Ack & Acknowledgment \\ \hline
AEED & Average End-to-End Delay \\ \hline
ALTN & Average Lifetime Network \\ \hline
CB & Capacity or Bandwidth \\ \hline
CBR & Constant Bit Rate \\ \hline
DAO & Destination Advertisement Object \\ \hline
DIO & DODAG Information Object \\ \hline
DIS & DODAG Information Solicitation \\ \hline
EC & Energy Consumption \\ \hline
ETX & Expected Transmission Count \\ \hline
IETF & Internet Engineering Task Force \\ \hline
IoT & Internet of Things \\ \hline
JFI & Jain Fairness Index \\ \hline
LALARPL & Learning Automata-based Load-Aware RPL \\ \hline
LLN & Low-Power and Lossy Network \\ \hline
LQI & Link Quality Indicator \\ \hline
L-RP & Linear Reward-Penalty \\ \hline
MAC & Medium Access Control \\ \hline
MP2P & Multipoint-to-Point \\ \hline
NL & Network Lifetime \\ \hline
OM & Other Metrics \\ \hline
P2MP & Point-to-Multipoint \\ \hline
P2P & Point-to-Point \\ \hline
PDR & Packet Delivery Ratio \\ \hline
PTR & Parent Trigger Reordering \\ \hline
RPL & Routing Protocol for Low-Power and Lossy Networks \\ \hline
TI & Traffic Index \\ \hline
TL & Traffic Load \\ \hline
UDP & User Datagram Protocol \\ \hline
\end{tabular}
\end{scriptsize}
\end{table}

\newpage

\begin{table}[H]
\caption{List of Parameters and their Definitions}
\label{tab:parameters}
\begin{scriptsize}
\centering
\begin{tabular}{|p{0.1\textwidth}|p{0.8\textwidth}|}
\hline
\textbf{Parameter} & \textbf{Definition} \\ \hline
$ALTN$ & Average Lifetime Network, a metric assessing the average operational lifetime of nodes in a network. \\ \hline
$C_i$ & Number of child nodes for node $i$, influencing the load handled by the node. \\ \hline
$CB_i$ & Capacity or bandwidth accessible to the parent node $i$. \\ \hline
$E_{\text{total}_i}$ & Total energy consumption for node $i$, considering power consumption in different operational states. \\ \hline
$F$ & Jain Fairness Index for throughput, measuring fairness in resource distribution among network nodes. \\ \hline
$g(x, \xi)$ & Function that modulates the influence of the maximum hop count dynamically. \\ \hline
$h(x, y, z)$ & Function that scales the squared difference between average and individual traffic indices. \\ \hline
$JFI_E$ & Jain's Fairness Index for energy consumption, measuring the fairness of energy usage across the network. \\ \hline
$LDI_i^p$ & Aggregated delay over each link for parent $p$, summing up all individual delays. \\ \hline
$LQI_{ij}$ & Link Quality Indicator between node $i$ and its $j$-th neighbor. \\ \hline
$M$ & Number of nodes still operational at the end of the simulation period. \\ \hline
$n$ & Number of nodes or actions, depending on the context in the formulas. \\ \hline
$N$ & Number of nodes in the network, or total number of actions, depending on context. \\ \hline
$NodeDelay_i$ & Total delay at node $i$, including processing, queuing, transmission, and propagation delays. \\ \hline
$numhop_i$ & Number of hops from node $i$ to the network root. \\ \hline
$p_i(n)$ & Probability of choosing action $i$ at time step $n$. \\ \hline
$P_i$ & Selection probability for the $i$-th parent node. \\ \hline
$P_I$ & Power consumption while idling. \\ \hline
$P_R$ & Power consumption while receiving. \\ \hline
$P_S$ & Power consumption while sleeping. \\ \hline
$P_T$ & Power consumption while transmitting. \\ \hline
$PS$ & Parent Set, a collection of potential parent nodes for a child node. \\ \hline
$R_{ij}$ & Data received from node $j$ during the time interval $\Delta t$. \\ \hline
$t_i$ & Time of death of the $i$-th node. \\ \hline
$t_I$ & Duration spent in idling state. \\ \hline
$t_R$ & Duration spent in receiving state. \\ \hline
$t_S$ & Duration spent in sleeping state. \\ \hline
$t_T$ & Duration spent in transmitting state. \\ \hline
$T_i$ & Traffic index associated with parent $i$. \\ \hline
$T_k$ & Aggregate traffic generated by node $k$. \\ \hline
$TI_i$ & Traffic Index of a designated parent node $i$. \\ \hline
$\alpha$ & Reward parameter that determines how strongly an action's probability increases upon receiving a reward. \\ \hline
$\beta$ & Penalty parameter that determines how strongly an action's probability decreases after being penalized. \\ \hline
$\delta$, $\gamma$, $\eta$, $\xi$ & Parameters used in network functions for hop count impact, traffic index scaling, and more. \\ \hline
$\Delta$ & Average end-to-end delay in the network. \\ \hline
$\Delta t$ & Time interval during which throughput is measured. \\ \hline
$f(x, y, z)$ & Function defining a damping factor that adjusts the impact of hop count differences. \\ \hline
$\lambda$ & Traffic rate, defined in packets per second. \\ \hline
$\phi_{\text{received}}$ & Total data packets successfully received. \\ \hline
$\phi_{\text{sent}}$ & Total data packets sent. \\ \hline
$\tau_i$ & Time the $i$-th packet travels from its source to its destination. \\ \hline
$\theta_i$ & Throughput for the $i$-th node. \\ \hline
$\wp$ & Predefined or estimated maximum lifetime for the surviving nodes. \\ \hline
$\zeta$ & Weighting parameter that balances hop count and traffic index in selection probability. \\ \hline
\end{tabular}
\end{scriptsize}
\end{table}

\section{RPL Challenges} \label{sec2}

\subsection{Congestion Problem in IoT}
Congestion occurs in a network when resource demand (e.g., bandwidth, routing capacity, buffer space) exceeds the available supply, leading to performance degradation \cite{anitha2023comprehensive_53}. In the context of networks, especially in IoT environments, congestion can lead to packet loss, increased latency, lower throughput, and inefficient energy use, which is particularly critical for battery-powered devices.
Congestion in networks signifies a state where network resources are overwhelmed due to excessive data packets sent through the network, leading to a bottleneck \cite{ShabaniBaghani2023_61, Mahyoub2021_62}. This situation results in a series of adverse effects, such as:

\begin{itemize}
    \item \textbf{Packet Loss:} When a network experiences congestion, the data packets arrive at a rate that exceeds the buffer’s capacity to store them for processing and forwarding. IoT devices, such as wireless sensor network nodes and LLN devices, have finite memory allocated for buffering packets. Once this memory (buffer) is filled due to congestion, any additional incoming packets cannot be accommodated, leading to what’s known as packet loss. This loss necessitates retransmissions if reliability is a requirement of the communication, which further contributes to the congestion, creating a feedback loop that exacerbates the network’s congested state.
    \item \textbf{Increased Latency:} Latency refers to the time a packet travels from its source to its destination. In congestion conditions, packets queue up at network nodes, waiting for their turn to be processed and forwarded. This queuing delay is a significant contributor to overall network latency. The more congested a network is, the longer the queues at network nodes are and, consequently, the higher the latency. For real-time applications and time-based monitoring, increased latency can severely degrade the quality of the service, leading to delays, destroying the cyber-physical system, or lags that impact user experience.
    \item \textbf{Lower Throughput:} Throughput in a network is the rate at which data is successfully delivered over a communication channel. Congestion leads to packet loss and increased latency, lowering the network's effective throughput. As packets are dropped or delayed, the data transmission rate effectively decreases. 

    \item \textbf{Energy Waste:} Congestion introduces significant energy waste in wireless sensor networks and IoT devices, which are often battery-powered and designed to operate efficiently to prolong battery life. When packets are lost due to congestion, they frequently need to be retransmitted to ensure the information reaches its intended destination. Each retransmission consumes energy for both the retransmitting device and potentially other devices in the network that participate in forwarding the packet. Moreover, devices in a congested network may need to stay in a higher power state longer to deal with the congestion, increasing energy consumption. This unnecessary energy expenditure reduces the overall lifetime of the devices and can be particularly problematic in IoT applications, where devices are expected to operate autonomously for extended periods \cite{lakshmi2024computational_11}.
\end{itemize}

\subsection{Congestion Control in RPL}

Congestion control in networking, especially within the context of the RPL and similar protocols used in IoT applications, is a critical process designed to maintain optimal network performance under varying load conditions. It encompasses a set of mechanisms and strategies to prevent or mitigate the negative effects of congestion—such as packet loss, increased latency, and decreased throughput—that occur when network traffic exceeds its capacity to process and forward data efficiently \cite{anitha2023comprehensive_53, jain2022_68}. Effective congestion control ensures that data flows smoothly through the network, optimizing resource utilization and enhancing the reliability and efficiency of data transmission, particularly in environments where network resources are constrained and must be judiciously managed to support the diverse needs of IoT applications \cite{Safaei2021_63}.

\begin{itemize}
    \item \textbf{Adaptive Retransmission Mechanisms:} To further mitigate congestion, RPL can incorporate adaptive retransmission mechanisms that adjust the rate at which data is sent and the criteria under which retransmissions occur. The network can reduce unnecessary data traffic by monitoring the success rates of packet deliveries and dynamically adjusting retransmission strategies. For instance, increasing the retransmission timeout could prevent exacerbating the congestion in conditions where packet loss is due to congestion rather than poor link quality.
    \item \textbf{Proactive Congestion Detection:} RPL can be enhanced with proactive congestion detection algorithms that identify potential congestion before it becomes problematic. The network can predict and address congestion risks by analyzing trends in data flow rates, queue lengths, and node capacities. Proactive measures might include rerouting traffic, adjusting transmission rates, or temporarily suspending non-critical data transmissions to allow the network to recover.
    \item \textbf{Multi-Path Routing:} To enhance load balancing and traffic distribution, RPL supports multi-path routing \cite{Taghizadeh2021_65}. This allows data to be sent over multiple paths, distributing the load evenly across the network and reducing the risk of any single path becoming a bottleneck. Multi-path routing contributes to congestion control and enhances network resilience by providing alternate routes for node or link failures.
    \item \textbf{Energy-Aware Congestion Control:} Given the energy constraints of devices in LLNs, RPL's congestion control mechanisms are designed to be energy-aware. This includes optimizing the trade-off between energy consumption and network performance. For example, while reducing the data transmission rate can alleviate congestion, it can also mean that devices spend more time in active communication states, consuming more energy \cite{venugopal2023_66}. RPL aims to balance these factors to maintain network efficiency without unduly depleting node batteries.
    \item \textbf{Integration with Application Layer Protocols:} Congestion control in RPL is not limited to the network layer but involves coordination with application layer protocols such as CoAP. By integrating congestion control strategies across layers, RPL ensures that application layer behaviours—such as data generation rates and priority data transmissions—are aligned with the network's congestion status. This cross-layer approach enables more sophisticated congestion management strategies that can adapt to the specific requirements and behaviours of IoT applications.
    \item \textbf{Community Engagement and Feedback Loops:} RPL incorporates mechanisms for community engagement and feedback loops, allowing the network to adapt to its nodes' collective behaviour. By aggregating feedback on congestion levels, packet loss rates, and throughput across different parts of the network, RPL can adjust its overall congestion control strategies to reflect the current network conditions. This collective intelligence approach ensures that RPL remains responsive to the dynamic nature of LLNs.
\end{itemize}

\subsection{Load Balancing in RPL}

In the expansive realm of the IoT, addressing load balancing in the RPL protocol presents a unique and pivotal strategy for orchestrating the distribution of computational and communication tasks across a vast network of heterogeneous devices. This orchestration transcends traditional network traffic management by optimizing the use of every iota of resource available, from bandwidth to processing power, thus ensuring that the network operates at its zenith of efficiency and reliability. The challenges stem from the inherent characteristics of IoT networks, such as resource constraints, dynamic network topologies, and the diverse requirements of IoT applications, necessitating meticulous management to conserve energy while fulfilling their roles \cite{rani2021rpl_17}.

The significance of load balancing in such a diversified ecosystem, facilitated by RPL protocol within IoT, cannot be overstated. The linchpin maintains equilibrium within the network, ensuring that no single node bears an undue burden that could precipitate its premature energy depletion or failure. By judiciously apportioning tasks, load balancing minimizes latency, optimizes energy use, and, by extension, amplifies the network's operational longevity and resilience. This harmonization of disparate device capabilities—ranging from robust servers that anchor the network to the myriad of energy-constrained sensors populating WSNs—is particularly crucial in applications where data integrity and timeliness are paramount, underscoring the indispensable role of load balancing in the fabric of IoT.

Critical challenges in implementing effective load balancing in the RPL protocol underscore the need for advanced strategies that address IoT devices and networks' unique demands and capacities. These efforts aim to create a seamless IoT ecosystem where devices can communicate and cooperate regardless of their underlying technology, ensuring a cohesive and efficient network operation \cite{darabkh2021rpl_01, tadigotla2020comprehensive_02, magubane2021extended_04, pancaroglu2021load_10, venugopal2023load_12}. Here are some of the critical challenges in implementing effective load balancing in RPL protocol:

\begin{itemize}
    \item \textbf{Resource Constraints:} IoT devices typically operate with limited computational power, memory, and energy supply. These constraints make it challenging to implement complex load-balancing algorithms that require significant processing power or memory resources. The challenge lies in designing lightweight load-balancing mechanisms that can operate efficiently within the limited capabilities of IoT devices \cite{Vaezian2022_58}.
    \item \textbf{Dynamic Network Topologies:} IoT networks are highly dynamic, with devices frequently joining or leaving the network \cite{Vaezian2022_58, Thiagarajan2023_56}. This dynamic nature results in fluctuating network topologies, complicating a balanced load distribution. Load-balancing solutions must be adaptive and capable of responding promptly to changes in the network structure without imposing excessive overhead.
    \item \textbf{Diverse Traffic Patterns:} IoT applications can generate highly diverse traffic patterns, ranging from periodic telemetry data transmissions to event-driven alerts. This diversity challenges load-balancing efforts, as algorithms must accommodate varying data rates, packet sizes, and transmission frequencies to prevent congestion and ensure fair resource allocation \cite{Mishra2022_59}.
    \item \textbf{Congestion Detection and Response:} Accurately detecting congestion in its early stages is critical for preemptive load balancing. However, due to IoT networks/RPL, traditional congestion detection mechanisms may not be directly applicable. Moreover, once congestion is detected, the protocol must swiftly reroute traffic without exacerbating the network's energy consumption or disrupting ongoing communications.
    \item \textbf{Energy-Efficiency Considerations:} In battery-powered IoT devices, energy conservation is paramount \cite{Grover2023_57}. Load balancing strategies must distribute network traffic evenly and minimize energy consumption. This requirement often leads to trade-offs between achieving optimal load distribution and prolonging the network's operational lifetime.
    \item \textbf{Compatibility and Standardization:} Ensuring that load balancing enhancements are compatible with the existing RPL specifications and interoperable across different implementations is crucial \cite{rani2021rpl_17}. Any modification to the RPL protocol to support load balancing must adhere to standardization efforts to facilitate widespread adoption and maintain network interoperability.
    \item \textbf{Qos Requirements:} IoT applications may have specific QoS requirements, such as low latency or high reliability \cite{rani2021rpl_17}. Load balancing mechanisms must meet these application-specific demands while managing network resources efficiently. Balancing the QoS requirements with the goal of load balancing adds another layer of complexity to the design and implementation of RPL enhancements.
\end{itemize}

To provide a solid foundation for understanding the evolution and enhancement of the RPL in the context of IoT environments, we have meticulously selected a collection of pivotal research papers from 2020 to 2024. The aim was to encapsulate the breadth of innovation and strategic advancements in RPL to address critical issues such as energy efficiency, load balancing, and network reliability. Our selection criteria were anchored in identifying works that proposed novel enhancements or modifications to RPL and demonstrated tangible improvements in network performance metrics such as packet delivery ratio, energy consumption, and network lifetime (Table \ref{tab:RPL2020to2024}). This was predicated on the understanding that the IoT paradigm is rapidly evolving, with escalating demands for more resilient, efficient, and scalable network routing protocols that cater to the burgeoning array of IoT devices and applications.

The ensuing comparison table serves as a panoramic snapshot, showcasing the diversity and ingenuity of approaches undertaken by researchers to fortify RPL against the multifaceted challenges intrinsic to LLNs. Through (Table 2\ref{tab:parameters-rpl}), readers can discern the trajectory of RPL enhancements over the years, where each entry highlights the primary aim, strategy, and strengths of the respective studies. This illuminates the progressive strides made in optimizing RPL for IoT applications and provides a clear vantage point from which to appreciate the nuanced strategies devised to strike a balance between competing objectives such as energy efficiency, network stability, and load distribution. By aggregating these insights, the table acts as a crucial reference point for stakeholders in the IoT ecosystem seeking to navigate the complexities of network routing in LLNs, offering a clear lens through which to gauge the state-of-the-art in RPL enhancements.

\newpage
\begin{scriptsize}
\begin{longtable}{|p{0.10\textwidth}|p{0.22\textwidth}|p{0.25\textwidth}|p{0.25\textwidth}|}
\caption{RPL Enhancements for Load Balancing (2020-2024)} \label{tab:RPL2020to2024} \\
\hline
\textbf{Year/RFC} & \textbf{Aim} & \textbf{Strategy} & \textbf{Strengths} \\ \hline
\endfirsthead
\caption{RPL Enhancements for Load Balancing (continued)} \\
\hline
\textbf{Year/RFC} & \textbf{Aim} & \textbf{Strategy} & \textbf{Strengths} \\ \hline
\endhead
2020 - \cite{haque2020energy_03} & Enhance RPL for IoT focusing on energy efficiency and reliability & Evaluate performance of ETX and Energy-based OFs; propose a hybrid OF & Identifies trade-offs between energy efficiency and reliability; proposes a balanced approach through a hybrid OF \\ \hline
2020 - \cite{kumar2020dcrl_19} & Address routing overhead, packet losses, and load imbalance in RPL-based IoT networks & Introduce DCRL-RPL framework with grid construction, ranking-based grid selection, and dual context-based OF selection & Demonstrate improved network lifetime, packet delivery ratio, and reduced routing overhead \\ \hline
2020 - \cite{seyfollahi2020lightweight_20} & Enhance RPL for IoT, focusing on load balancing & Introduce L2RMR with a novel OF and PF to prevent HDP, optimizing path length and load distribution & Significantly reduces packet loss, delay, and energy consumption, outperforming traditional RPL \\ \hline
2020 - \cite{pereira2020increased_22} & Improve load balancing and extend network lifetime in IoT & Introduce NIAP metric for balancing energy consumption, relying on average power estimation & Increases network lifetime by up to 24\%, improves packet delivery ratio and reduces delay \\ \hline
2020 - \cite{wang2020sl_23} & Address instability and inefficiency in RPL's load balancing for IoT & Propose SL-RPL with stability-aware mechanism, utilizing PTR and ETX for parent selection & Enhances network stability and performance, reducing parent changes, packet loss, and energy usage \\ \hline
2020 - \cite{sebastian2020child_36} & Address load imbalance in RPL for IoT & Introduce Ch-LBRPL to improve load balance using a child count method, reducing parent switching and enhancing energy efficiency & More effective at achieving load balance, improving network stability and energy consumption \\ \hline
2020 - \cite{rana2020ebof_38} & Tackle load imbalance in RPL for IoT & Introduce EBOF combining ETX and CC for optimal path selection, extending network lifetime & Enhances network performance by balancing energy consumption and prolonging operational sustainability \\ \hline
2020 - \cite{stoyanov2020comparative_25} & Evaluate QU-RPL's load-balancing in RPL for IoT & Comparative analysis of RPL and QU-RPL focusing on power consumption, PDR, and latency & Finds QU-RPL does not significantly improve over traditional RPL, suggesting a need for further development \\ \hline
2020 - \cite{vaziri2020brad_43} & Enhance energy-aware parent selection and congestion avoidance in RPL for IoT & Propose Brad-OF using ETX, delay, and residual energy for parent selection and a metric for congestion avoidance & Increases network lifetime by up to 65\% and reduces packet loss by up to 81\% \\ \hline
2020 - \cite{mahyoub2020efficient_47} & Address N2N communication inefficiencies in LLNs for IoT & Propose HRPL, integrating link-state routing with RPL for efficient N2N routes and employing adaptive reporting and SSSP mechanisms & Significantly improves packet delivery ratio, reduces delay and energy consumption, maintaining RPL compatibility \\ \hline
2020 - \cite{sankar2020ct_49} & Extend network lifespan and reduce data traffic in IoT & Introduce CT-RPL with cluster formation, CH selection, and route establishment based on RER, QU, and ETX & Enhances network lifetime by 30-40\% and packet delivery ratio by 5-10\% \\ \hline
2021 - \cite{sirwan2021adaptive_50} & Facilitate load-efficient IoT connectivity with anticipated device number surge & Leverage self-coordinating networks and distributed learning for dynamic communication parameter adaptation & Demonstrate improvements in reliability and traffic efficiency with lightweight learning \\ \hline
2021 - \cite{magubane2021extended_04} & Improve RPL in IoT by incorporating buffer occupancy for load balancing & Introduce ECLRPL, using a buffer occupancy metric in routing decisions to enhance throughput and network lifetime & Significantly outperforms standard RPL and CLRPL in packet delivery, power efficiency, and network delay \\ \hline
2021 - \cite{zheng2021load_06} & Address load imbalance in LLNs with RPL by proposing a clustering-based protocol & Use non-uniform clustering and cluster head rotation based on node energy and priority for balanced load & Enhances network stability and efficiency by achieving balanced traffic distribution \\ \hline
2021 - \cite{idrees2021energy_16} & Develop an energy-efficient, load-balanced routing protocol for IoT networks & Incorporate a novel parent selection algorithm in EL-RPL, considering energy and packet counts & Outperforms existing protocols in energy conservation, control packet reduction, and extending network lifetime \\ \hline
2021 - \cite{acevedo2021wrf_30} & Enhance load balancing in high-traffic sensor networks & Introduce WRF-RPL with a routing metric considering remaining energy and parent count & Outperforms standard RPL in network lifetime, packet delivery, and energy consumption \\ \hline
2021 - \cite{fatemifar2021new_18} & Improve routing in IoT networks & Propose C-Balance with a dual-ranking system for cluster formation and routing, using ETX, hop count, and energy metrics & Improves network longevity and energy efficiency, though increases end-to-end delay \\ \hline
2021 - \cite{royaee2021designing_21} & Address load imbalance in RPL for IoT & Develop AMRRPL with ant colony optimization for rank computation and stochastic automata for parent selection & Demonstrate improvements in packet delivery, network lifetime, energy efficiency, and convergence \\ \hline
2021 - \cite{yassien2021routing_24} & Address load balancing challenges in RPL for IoT & Introduce LBTB, combining neighbour count and node power with a modified trickle timer for message distribution & Reducing convergence time by up to 68\%, power consumption by 16\%, and delay by 56\% \\ \hline
2021 - \cite{arunachalam2021load_26} & Mitigate hotspot problem and improve data aggregation in IoT with RPL & Propose LoB-RPL with a composite metric for parent selection and adaptive trickle parameters & Significantly improves packet delivery, network lifetime, energy efficiency, and control overhead reduction \\ \hline
2021 - \cite{zarzoor2021optimizing_45} & Optimize RPL performance in IoT for reducing node congestion and latency & Introduce E-MHOF with a three-layer approach for parent and path selection, and child node minimization & Demonstrates significant improvements in network lifetime and latency reduction \\ \hline
2021 - \cite{abdullah2021efficient_32} & Improve routing and address node unreachability in LLNs for IoT & Propose MSLBOF with Memory Utilization metrics for sink selection and load balancing & Significantly reduces packet loss and improves network stability compared to standard MRHOF \\ \hline
2022 - \cite{hadaya2022new_33} & Address energy consumption and inefficiency in RPL for IoT & Propose a novel RPL OF incorporating Load, Residual Energy, and ETX to enhance network lifetime and efficiency & Shows a PDR increase of 58.425\%, a decrease in packet loss ratio, and a reduction in power consumption \\ \hline
2022 - \cite{anita2022learning_35} & Optimize RPL for energy efficiency and load balancing in IoT & Introduce a methodology using learning automata and lexical composition for critical routing metrics selection & Significantly improves packet delivery ratio, energy consumption, and network stability \\ \hline
2022 - \cite{Awiphan2022_42} & Improve load balancing in RPL-based by reducing node overload. & Identify neighbours at the same rank and exchange metrics like available connections and ETX to better select network parents. & Improved packet delivery and reduced packet loss compared to traditional methods, optimizing network traffic distribution. \\ \hline
2022 - \cite{koosha2022fahp_05} & Enhance routing in RPL-based IoT networks & Develop FAHP-OF using Fuzzy Logic and AHP for dynamic parent selection optimization & Improves E2ED and PDR, enhancing network reliability and efficiency \\ \hline
2022 - \cite{kaviani2022cqarpl_48} & Propose CQARPL for IoT applications under heavy traffic conditions & Incorporates congestion control and enhanced QoS into RPL; uses multiple metrics for routing decisions & Enhances network lifetime, reduces queue loss ratio, improves packet reception, and lowers delay \\ \hline
2023 - \cite{jagirhussain2023berpl_27} & Introduce BE-RPL to address mobility issues in IoT LLN & Enhances RPL with mobility awareness and energy efficiency; focuses on load balancing and reactive parent selection & Demonstrates improvements in energy utilization, network control overhead, and packet delivery ratio \\ \hline
2023 - \cite{kalantar2023energy_29} & Tackle energy management and traffic balance in IoT networks & Introduces ELBRP with ECAOF for parent node selection based on energy and congestion & Shows significant advancements in energy efficiency and packet delivery, with a slight increase in control overheads \\ \hline
2023 - \cite{subramani2023weighted_31} & Achieve load balance and efficient routing in IoT networks & Propose WSM-OF using a combination of ETX, LQL, RE, and Child Count & Improves control overhead, jitter, packet delivery ratio, energy consumption, and network lifetime by up to 7.8\% \\ \hline
2023 - \cite{tiwari2023improved_39} & Enhance RPL for WSNs with integrated mobility management & Focus on micro-mobility to optimize energy consumption and load balancing & Reduces energy consumption, enhances packet delivery ratios, and ensures stable network operation \\ \hline
2023 - \cite{venugopal2023congestion_40} & Address load balancing and congestion in LLNs for IoT & Introduce CEA-RPL with CEA-OF leveraging Queue Occupancy, Expected Lifetime, and Child Count & Enhances power consumption, packet receiving rate, end-to-end delay, and network lifetime \\ \hline
2023 - \cite{ahmed2023tfuzzy_44} & Address load balancing in RPL for IoT networks & Propose TFUZZY-OF integrating fuzzy logic with TOPSIS & Enhances PDR and reduces E2ED compared to traditional methods \\ \hline
2024 - \cite{lei2024reinforcement_41} & Optimize RPL routing in IoT environments using DRL & Develop RARL with a DRL model for intelligent routing decisions & Outperforms existing methods in network lifetime, queue loss ratio, packet reception ratio, and delay \\ \hline
2024 - \cite{tabouche2024tlr_46} & Address load-balancing issues in IIoT networks over 6TiSCH & Develop TLR with a traffic-aware proactive path selection strategy & Demonstrates superiority in throughput, reliability, latency, and energy efficiency over conventional RPL \\ \hline
2024 - \cite{alilou2024qfs_51} & Integrate Q-learning and FSR in RPL for IoT to enhance mobility and energy efficiency & Propose QFS-RPL for efficient load balancing and improved PDR & Shows superior performance, especially in mobile node environments, enhancing network throughput and lifetime \\ \hline
2024 - \cite{shashidhar2024adaptive_52} & Enhance multimedia data transmission efficiency in IoT networks & Introduce ARPLO with a grid-based structure and ADNN for data classification & Improve energy efficiency, throughput, PDR, and network lifespan while reducing control overhead and delay \\ \hline
\end{longtable}
\end{scriptsize}

\newpage

\begin{figure}
    \centering
    \includegraphics[width=1.0\linewidth]{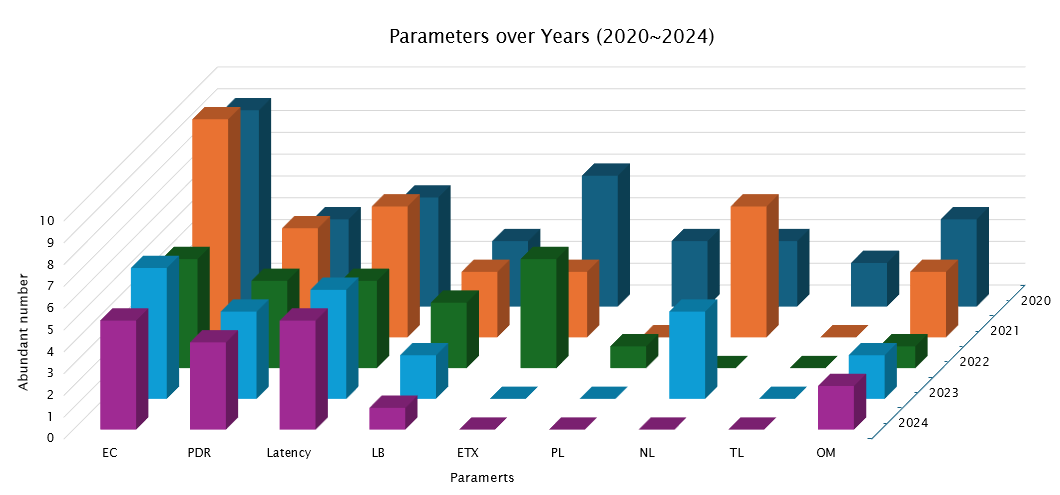}
    \caption{Parameters occurrence over the years 2020 - 2024 *}
    \label{fig:Year-parameters}
\end{figure}

\begin{scriptsize}
\begin{longtable}[c]{| p{1cm} | p{0.5cm} | p{13cm} |} 
\caption{Key parameters in recent research on RPL's Load-balancing} \\
\hline
\textbf{Year} & \textbf{Ref} & \textbf{Key Parameters} \\
\hline
\endhead
\multirow{11}{*}{2020} & \cite{haque2020energy_03} & ETX, Energy \\ \cline{2-3}
 & \cite{kumar2020dcrl_19} & EC, LB, Overhead, PDR \\ \cline{2-3}
 & \cite{seyfollahi2020lightweight_20} & LB, Path Length, PL, Latency \\ \cline{2-3}
 & \cite{pereira2020increased_22} & EC, PDR, Latency \\ \cline{2-3}
 & \cite{wang2020sl_23} & Stability (PTR, ETX), PL, EC \\ \cline{2-3}
 & \cite{sebastian2020child_36} & Child Count, EC, Overhead \\ \cline{2-3}
 & \cite{rana2020ebof_38} & EC, ETX, Child Count, NL \\ \cline{2-3}
 & \cite{stoyanov2020comparative_25} & EC, PDR, Latency \\ \cline{2-3}
 & \cite{vaziri2020brad_43} & ETX, Latency, EC, TL, PL \\ \cline{2-3}
 & \cite{mahyoub2020efficient_47} & Link-state, PDR, MAC state, Latency, EC \\ \cline{2-3}
 & \cite{sankar2020ct_49} & EC, Queue, ETX, NL, TL \\ \hline
\multirow{11}{*}{2021} & \cite{sirwan2021adaptive_50} & Reliability, Communication Efficiency \\ \cline{2-3}
 & \cite{magubane2021extended_04} & Buffer Occupancy, PDR, EC, Latency \\ \cline{2-3}
 & \cite{zheng2021load_06} & Clustering, Stability, TL \\ \cline{2-3}
 & \cite{idrees2021energy_16} & EC, NL \\ \cline{2-3}
 & \cite{acevedo2021wrf_30} & EC, LB, Parent Node Count \\ \cline{2-3}
 & \cite{fatemifar2021new_18} & ETX, Hop Count, EC, Number of Node Children, Network Longevity \\ \cline{2-3}
 & \cite{royaee2021designing_21} & Congestion Mitigation, NL, EC, PDR \\ \cline{2-3}
 & \cite{yassien2021routing_24} & Neighbour Count, EC, Trickle Timer, Convergence Time, Latency \\ \cline{2-3}
 & \cite{arunachalam2021load_26} & Composite Metric, Trickle Timer, PDR, NL \\ \cline{2-3}
 & \cite{zarzoor2021optimizing_45} & Congestion Mitigation, ETX, RSSI, EC, Latency \\ \cline{2-3}
 & \cite{abdullah2021efficient_32} & Memory Utilization, LB (Multi-Sink), PL \\ \hline
\multirow{7}{*}{2022} & \cite{hadaya2022new_33} & LB, EC, ETX, NL, PDR \\ \cline{2-3}
 & \cite{anita2022learning_35} & EC, LB, Hop Count, ETX, TL, PDR \\ \cline{2-3}
 & \cite{Awiphan2022_42} & ETX, PDR, Overhead \\ \cline{2-3}
 & \cite{koosha2022fahp_05} & Hop-count, ETX, RSSI, PDR, Latency \\ \cline{2-3}
 & \cite{kaviani2022cqarpl_48} & Congestion, QoS, ETX, Hop Count, NL, QLR, PRR, Latency \\ \hline
\multirow{7}{*}{2023} & \cite{jagirhussain2023berpl_27} & Mobility Management, EC, LB, PDR \\ \cline{2-3}
 & \cite{kalantar2023energy_29} & EC, Congestion, NL, Latency, Overhead \\ \cline{2-3}
 & \cite{subramani2023weighted_31} & ETX, Link Quality, EC, Child Count, Jitter, Parent Switching, Latency, NL \\ \cline{2-3}
 & \cite{tiwari2023improved_39} & Mobility Management, EC, PDR, Network Stability \\ \cline{2-3}
 & \cite{venugopal2023congestion_40} & Congestion, EC, Queue, NL, Latency, PDR \\ \cline{2-3}
 & \cite{ahmed2023tfuzzy_44} & Hop Count, ETX, RSSI, PDR, Latency \\ \hline
\multirow{4}{*}{2024} & \cite{lei2024reinforcement_41} & EC, NL, Queue \\ \cline{2-3}
 & \cite{tabouche2024tlr_46} & TL, Queue, Throughput, Latency, EC \\ \cline{2-3}
 & \cite{alilou2024qfs_51} & Overhead, PDR, Latency, Throughput \\ \cline{2-3}
 & \cite{shashidhar2024adaptive_52} & EC, Throughput, PDR, Overhead, Latency \\ \hline
\end{longtable} \label{tab:parameters-rpl}
\begin{scriptsize}
*
Energy Consumption: EC |
Packet Delivery Ratio: PDR |  
Latency: Average End-to-End Delay |
Load balancing: LB |
Expected Transmission Count: ETX |
Packet Loss: PL |
Network Lifetime: NL |
Traffic Load: TL |
Other Metrics: OM
\end{scriptsize}
\end{scriptsize}

\clearpage

\section{Learning Automata}\label{sec3}

\begin{figure}[h]
    \centering
    \includegraphics[width=0.4\linewidth]{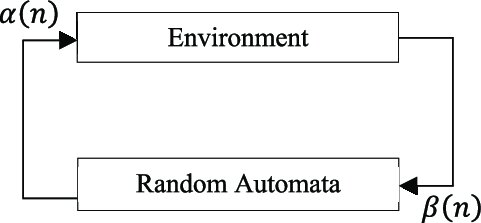}
    \caption{Stochastic Learning Automata \cite{Kordestani2021advances}}
    \label{fig:LA}
\end{figure}

A Learning Automaton (LA) is a model for decision-making in stochastic environments that dynamically adjusts its strategies based on feedback to optimize overall performance. This document outlines Learning Automata's operational principles and implementation details, focusing on the Linear Reward-Penalty (L\_R-P) update scheme \cite{Kordestani2021advances, Homaei2019}. Below are key definitions related to the components of a Learning Automaton (see Figure \ref{fig:LA}):

\begin{itemize}
    \item \textbf{Environment}: The dynamic setting in which the automaton operates. It is a source of stimuli where the automaton performs actions and receives feedback based on the outcomes of these actions.
    \item \textbf{Actions}: These are the decisions or moves the automaton can make. The probability of choosing each action is updated continuously based on the feedback received from the environment.
    \item \textbf{Response}: The feedback from the environment following an action, which can be positive (reward) or negative (penalty), influencing how the probabilities of actions are updated.
    \item \textbf{Learning Algorithm}: The method used to update the probabilities of actions in response to the feedback, to enhance the efficiency and effectiveness of the decision-making process.
\end{itemize}

\subsection{Pseudocode for Learning Automaton}
The pseudocode below demonstrates the implementation of a Learning Automaton using the Linear Reward-Penalty (L\_R-P) update scheme. This scheme adjusts the probability of selecting actions based on the feedback received, optimizing the automaton's responses to environmental changes.

\begin{algorithm}
\caption{Learning Automaton Procedure}
\begin{algorithmic}[1]
\State \textbf{Input:} Environment
\State \textbf{Output:} Updated probability vector $P$
\State Initialize $P$ with equal probabilities for each action
\While{not terminate\_condition}
    \State Select an action based on the probability distribution $P$
    \State Perform the action in the environment
    \State Receive feedback (reward or penalty) from the environment
    \If{feedback is reward}
        \State $P[\text{selected\_action}] \gets P[\text{selected\_action}] + \alpha \cdot (1 - P[\text{selected\_action}])$
        \For{each action $j \neq \text{selected\_action}$}
            \State $P[j] \gets (1 - \alpha) \cdot P[j]$
        \EndFor
    \ElsIf{feedback is penalty}
        \State $P[\text{selected\_action}] \gets (1 - \beta) \cdot P[\text{selected\_action}]$
        \For{each action $j \neq \text{selected\_action}$}
            \State $P[j] \gets \frac{\beta}{N - 1} + (1 - \beta) \cdot P[j]$
        \EndFor
    \EndIf
    \State Check the terminate\_condition
\EndWhile
\State \Return $P$
\end{algorithmic}
\end{algorithm}

\clearpage

\clearpage

\subsection{Linear Reward-Penalty Scheme}
The Linear Reward-Penalty (LRP) scheme described here is a method used in reinforcement learning to adjust the probabilities of selecting certain actions based on the feedback (reward or penalty) received from the environment \cite{Rezvanian2018, Rezvanian2019}. This type of scheme is especially relevant in contexts where decisions are probabilistic and learning from the outcomes is necessary to improve performance over time. Here's a detailed breakdown of the scheme:

\begin{itemize}
    \item \textbf{Updating Probabilities When an Action is Rewarded:}

When action \( \alpha_i \) is rewarded, the probability of choosing this action in the next time step, \( p_i(n+1) \), is increased. The formula used is (Equation ~\ref{eq:reward_update}):
\begin{equation}
p_{i}(n+1) = p_{i}(n) + \alpha (1 - p_{i}(n)) \label{eq:reward_update}
\end{equation}
This equation means that the new probability \( p_i(n+1) \) is the old probability \( p_i(n) \) increased by a fraction \( \alpha \) of the remaining probability space \( 1 - p_i(n) \). The parameter \( \alpha \) controls how much the probability increases — larger \( \alpha \) results in a larger increase.

For all other actions \( j \neq i \), the probability of selecting each of these actions is reduced proportionally (Equation ~\ref{eq:reward_update_others}):
\begin{equation}
p_{j}(n+1) = (1 - \alpha)p_{j}(n) \label{eq:reward_update_others}
\end{equation}
Here, each other action's probability is scaled down by the factor \( 1 - \alpha \), ensuring that the total probability across all actions remains equal to 1.

    \item \textbf{Updating Probabilities When an Action is Penalized:}

Conversely, if action \( \alpha_i \) is penalized, its probability is reduced (Equation ~\ref{eq:penalty_update}):
\begin{equation}
p_{i}(n+1) = (1 - \beta) p_{i}(n) \label{eq:penalty_update}
\end{equation}
In this equation, \( \beta \) is the penalty parameter. The new probability \( p_i(n+1) \) is the previous probability \( p_i(n) \) reduced by a factor of \( \beta \), meaning a larger \( \beta \) decreases the probability more significantly.

For all other actions \( j \neq i \), their probabilities are updated as follows (Equation ~\ref{eq:penalty_update_others}):
\begin{equation}
p_{j}(n+1) = \frac{\beta}{r - 1} + (1 - \beta)p_{j}(n) \label{eq:penalty_update_others}
\end{equation}
This adjustment ensures that the total probability is still 1. The increase for each other action is partly a fixed amount \( \frac{\beta}{r - 1} \), which redistributes the reduced probability of the penalized action equally among them, and partly the old probability scaled down by \( 1 - \beta \).

\item \textbf{Parameters and their Roles:}

\begin{itemize}
    \item \( \alpha \) (reward parameter): Determines how strongly an action's probability is increased upon receiving a reward.
    \item \( \beta \) (penalty parameter): Determines how strongly an action's probability decreases after being penalized.
    \item \( r \) (total number of actions): Affects the redistribution of probabilities when an action is penalized.
\end{itemize}

The LRP scheme is a straightforward yet effective adaptive strategy for balancing exploration (trying out different actions) and exploitation (favoring actions that have previously led to positive outcomes) in environments where actions have probabilistic outcomes. This method ensures that actions that lead to success are more likely to be chosen in the future, while those that lead to negative outcomes are less likely to be repeated.
\end{itemize}

\section{Proposed LALARPL}\label{sec4}

This section comprehensively elucidates the Learning Automata-based load-aware RPL (LALARPL), a distributed load-balancing algorithm tailored for the RPL. Aimed to enhance network efficiency and reliability significantly, LALARPL dynamically distributes traffic loads across pathways within LLN environments. Central to the LALARPL framework is the Traffic Index (TI) parameter, which was devised to quantify the burden of traffic load accurately. The TI of a designated parent node \(i\) is mathematically expressed as (Equation~\ref{eq:TI}):

\begin{equation}
TI_{i} = \frac{\sum_{k \in \mathbb{N}} \theta_{k_{i}} \times T_{k}}{CB_{i}} \label{eq:TI}
\end{equation}

Herein, \(TI_i\) represents the traffic index pertinent to a specific parent node \(i\). The coefficient \(\theta_{k_{i}}\) signifies the traffic contribution from a child node \(k\) routed via the parent node \(i\). The term \(T_k\) denotes the aggregate traffic generated by node \(k\), and \(CB_i\) encapsulates the capacity or bandwidth accessible to the parent node \(i\). Notably, TI values are restricted to a range between zero and one, with one value indicating full utilization of a parent node's capacity. LALARPL's methodology is bifurcated into two key phases: Parent Set Formation and Load Balancing:

\begin{itemize}
    \item \textbf{Parent Set Formation:} Initially, a collection of proximate parents is formulated for every child node. This phase solely utilizes DIO-indicator packets, distinguishing it from subsequent procedures. During this phase, parent sets are established for each node through a process where parent nodes broadcast DIO-indicator packets within their communicable range. These packets are characterized by three specific fields: the IP address of the broadcasting node, the minimal number of hops to the root from a parent node, and the parent node's traffic index. Prior to the transmission of a DIO-indicator packet, the aforementioned fields are initialized by the parent node. Upon receipt of DIO-indicator packets, other nodes in the network undertake the update of their parent tables, adjusting the sending priorities based on the data received within their respective parent sets as delineated below:

\begin{itemize}
    \item Should a node receive merely a singular DIO-indicator packet, the information contained within is recorded in its parent set table, and the originating node is incorporated into its parent set with a selection probability assigned as 1.
    
    \item Conversely, in scenarios where multiple DIO-indicator packets are received, nodes are tasked with selecting between a minimum of two and a maximum of five parent nodes. The selection criteria encompass proximity (regarding the number of hops) and the residual energy of the potential parent nodes. These chosen nodes are added to the recipient node's parent set or routing path. The formulation for the selection probability of each potential parent node is given as follows (Equation ~\ref{eq:prob_calc}):
    
\begin{equation}
\begin{aligned}
\forall i \leq N \quad P_i = & \; \zeta \left( \frac{\frac{1}{\text{numhop}_i}}{\sum_{j=1}^{N} \frac{1}{\text{numhop}_j}} \right) & + (1 - \zeta) \left( \frac{T_i}{\sum_{j=1}^{N} T_j} \right)
\end{aligned}
\label{eq:prob_calc}
\end{equation}

Herein:
\begin{itemize}
    \item \(i\) indexes the enumeration of parent senders,
    \item \(T_i\) quantifies the traffic index associated with parent \(i\),
    \item \(\text{numhop}_i\) specifies the number of hops from node \(i\) to the network root,
    \item \(N\) represents the total count of potential parent nodes under consideration,
    \item \(\zeta\) is a weighting parameter within the interval [0,1], modulating the relative contributions of hop count and traffic index to the selection probability.
\end{itemize}

\end{itemize}

\begin{algorithm} [ht]
\caption{Load Balancing Algorithm based on Learning Automata}
\begin{algorithmic}[1]

\For{each Parent $P_i$}
    \State $C_i \gets \{\}$ \Comment{ParentSet of $P_i$, which is empty}
    \If{there is one P}
        \State Add $P_1$ to $PS_i$
    \Else
        \State Add $P_i$ to $d_i$, where $\min \{ \text{distance} (n_i, P_i)\}$
    \EndIf
\EndFor

\For{each Parent Set $PS_i$}
    \State generate (Data packet)
    \State select one P in $PS_i$; \Comment{using Learning Automata}
    \State send (Data packet) to the selected P
    \State waits for ACK
    \If{received (Data packet)}
        \State send (ACK Packet) to the sender node
    \EndIf
    \If{received (ACK packet)}
        \If{ACK == “reward”} \Comment{P with low load}
            \State Reward the P
        \ElsIf{ACK == “penalty”} \Comment{P with high load}
            \State Penalize the P
        \EndIf
    \EndIf
\EndFor

\end{algorithmic}
\end{algorithm}
   
\item \textbf{Distributed Load Balancing:} Following the parent set formation stage, the procedure for data transmission commences. Child nodes initiate the transmission of data packets toward their selected parent nodes, as indicated within their parent set tables. Concomitantly, parent nodes respond by dispatching Acknowledgment (Ack) packets, which encapsulate the sender's identification number and the traffic index. A parameter designated as \(p\) is introduced to curtail the message volume and diminish the network load. This parameter dictates that each node transmits a set of \(p\) data packets to a predetermined parent node, which issues a single Ack packet in response to the accumulation of \(p\) packets.
Additionally, the architecture of the routing tables is elucidated, revealing four principal fields: the Parent IP address, the selection probability assigned to that parent, the Traffic Index, and the hop count to the network root. An integral feature of the algorithm is the incorporation of a learning automaton within each child node. This automaton executes operations correlating with the number of parent nodes listed in the node's routing table or parent set, thereby enabling adaptive decision-making based on dynamic network conditions. The methodology governing the issuance of rewards or penalties after the receipt of an Ack packet is articulated as follows:

\begin{itemize}
    \item A reward is allocated when the parent's traffic index (the Ack's sender) is observed to be less than 50\% of the average traffic index pertinent to other parents within the identical set.
    
    \item Should the traffic index exceed 50\% yet remain below 80\% of the average traffic index of the other parents in the set, and the parent node exhibits the minimal number of hops to the root, a reward is similarly conferred.
    
    \item Conversely, a penalty is imposed when the traffic index surpasses the average traffic index associated with other parents in the set.
\end{itemize}

The formulations for rewards and penalties are articulated as follows (Equations \ref{eq:alpha_calc} and \ref{eq:beta_calc}):
\begin{equation}
\alpha = \alpha_1 + \frac{\delta \cdot TI_i + f(\text{max hop}, \text{num hop}_i, \gamma)}{\delta \cdot \max(TI) + g(\text{max hop}, \xi)} + c_1
\label{eq:alpha_calc}
\end{equation}

\begin{equation}
\beta = \alpha_2 + \frac{\delta \cdot h(\text{avg } TI, TI_i, \eta) + \text{num hop}_i}{\delta \cdot \text{avg } TI + g(\text{max hop}, \xi)} + c_2
\label{eq:beta_calc}
\end{equation}

Where:
\begin{itemize}
\item \( f(x, y, z) = x - e^{\gamma \cdot y} \) with \( \gamma \) being a damping factor that adjusts the impact of hop count differences.
\item \( g(x, \xi) = \xi \cdot \ln(x + 1) \) where \( \xi \) helps modulate the influence of the maximum hop count dynamically.
\item \( h(x, y, z) = z \cdot (x - y)^2 \) where \( \eta \) scales the squared difference between average and individual traffic indices, emphasizing deviations.
\item \( \max(TI) \) and \( \text{avg } TI \) are computed as the maximum and average traffic indices among the set, potentially including more sophisticated aggregation rules based on network topology or traffic patterns.
\end{itemize}
 In simulations of network traffic, it was observed that increasing the hop count linearly increases the network delay under low traffic conditions. However, as traffic intensifies, the delay increases exponentially. The function \( f(\text{max hop}, \text{num hop}_i, \gamma) \) was calibrated with real data to model this behaviour accurately, significantly improving the predictive performance of the model in high-traffic scenarios.
\end{itemize}

\section{Simulation Results}\label{sec5}
In the conducted study, the performance of the proposed Learning Automata-based Load-Aware Routing Protocol for LLNs (LALARPL) was meticulously evaluated and then compared against seven other protocols specifically designed for IoT/LLNs, denoted as Protocols ECLRPL~\cite{magubane2021extended_04}, FAHP-OF~\cite{koosha2022fahp_05}, NUCRPL~\cite{zheng2021load_06}, WRF-RPL~\cite{acevedo2021wrf_30}, TLR 46~\cite{tabouche2024tlr_46}, LBS-RPL~\cite{Awiphan2022_42}, and CEA-RPL~\cite{venugopal2023congestion_40}. Using the NS-2 discrete event simulator, a detailed assessment was conducted within a simulated environment designed to emulate potential real-world IoT deployment scenarios closely. This environment, characterized by a \(1000 \times 1000\) meter area populated under three different scenarios with 50, 100, and 150 static nodes, provided a comprehensive backdrop for examining the operational performance of each protocol, with the simulation parameters elaborately detailed in Table~\ref{tab:Simulation-parameters-extended}.

Crucial performance metrics, including Jain's Fairness Index in Throughput, were utilized to assess the equitable allocation of network resources across all nodes---a fundamental aspect to ensure fair bandwidth distribution within the IoT ecosystem. Additionally, the Packet Delivery Ratio (PDR) and Latency were thoroughly investigated as key indicators of each protocol's reliability and efficiency. They measured the success rate of packet deliveries and the time required for packets to traverse from their origin to their intended destinations, respectively. The inclusion of control message sizes for DIO, DAO, DIS, and DAO-Ack, as well as traffic rates defined by \(\lambda = 0.1\) packets per second and \(\lambda = 0.2\) packets per second, further enriched the simulation's scope. This methodological approach highlighted the comparative advantages and potential drawbacks of LALARPL relative to the other considered protocols and provided insightful revelations about the adaptability and operational effectiveness of these routing enhancements in anticipated IoT applications. Such discoveries emphasize the significant impact these protocols can have in shaping network environments to be more resilient, efficient, and fair.
\newpage
\begin{table}[t]
\centering
\begin{scriptsize}
\caption{SIMULATION PARAMETERS}
\label{tab:Simulation-parameters-extended}
\begin{tabular}{ll}
\toprule
\textbf{Parameter} & \textbf{Value} \\
\midrule
Simulator & NS-2.34 \\
Traffic Type & Constant Bit Rate (CBR) over UDP \\
Simulation Area & \(1000 \, \text{m} \times 1000 \, \text{m}\) \\
Simulation Time & 1000 s \\
Number of Nodes & 50, 100, 150 \\
Sink Placement & Centralized \\
Node Placement & Random \\
Topology & RPL tree-based \\
MAC Layer Protocol & IEEE 802.15.4 \\
Data Rate & 250 kbps \\
Bandwidth & Up to 250 kbps (consistent with IEEE 802.15.4) \\
Radio Range & 100 m \\
Packet Size & 50 bytes (maximum for IEEE 802.15.4) \\
Energy Model & Enabled \\
Initial Energy per Node & 2 Joules \\
Mobility Model & Static nodes \\
Routing Protocol & LALARPL, and others for comparison \\
DIO Message Size & 80 bytes \\
DAO Message Size & 100 bytes \\
DIS Message Size & 77 bytes \\
DAO-Ack Message Size & 80 bytes \\
Traffic Rate (\(\lambda\)) & 0.1, 0.2 packets/s \\
\bottomrule
\end{tabular}
\end{scriptsize}
\end{table}

\subsection{Packet Delivery Ratio Test}

Various factors influence the packet delivery rate in the RPL tree structure. Significant challenges such as link failure, congestion rates, and potential collisions within the network are recognized. These challenges can substantially impact the packet delivery rate. In addition, the PDR is defined as follows (Equation ~\ref{eq:pdr-formula}):

\begin{equation}
PDR = \frac{\sum_{i=1}^n \phi_{\text{received}, i}}{\sum_{i=1}^n \phi_{\text{sent}, i}}
\label{eq:pdr-formula}
\end{equation}

where:
\begin{itemize}
  \item \( \phi_{\text{received}} \) denotes the total data packets successfully received,
  \item \( \phi_{\text{sent}} \) represents the total data packets sent.
\end{itemize}

The Packet Delivery Ratio (PDR) is a critical metric in network performance, reflecting the reliability and efficiency of the routing protocol. The LALARPL method showcases significant advancements in this domain by integrating a graph degree restriction for optimized graph formation and multipathing mechanisms for alternate parent selection. These strategies and the implementation of distributed learning automata to weigh the value of parents dynamically have markedly enhanced PDR across various network sizes and packet arrival rates.

In a comparative analysis, the LALARPL method has outperformed its counterparts in all tested scenarios. Specifically, for a network of 50 nodes with a packet arrival rate ($\lambda$) of 0.1 packets per second, LALARPL recorded a PDR of 0.96, demonstrating an improvement of $1.06\%$ over the nearest competing protocol, TLR (PDR of 0.95). At the same network size but with a higher packet arrival rate ($\lambda = 0.2$ packets per second), LALARPL's PDR of 0.91 is $3.41\%$ better than the next best protocol, LBS-RPL (PDR of 0.9). Expanding the network to 100 nodes, LALARPL's superiority remains evident. With a $\lambda$ of 0.1, LALARPL achieved a PDR of 0.93, which is $6.45\%$ higher than the other methods' average PDR (0.87). At a $\lambda$ of 0.2, LALARPL's PDR of 0.9 is $8.43\%$ better than the competing protocols' average PDR (0.83). In denser networks of 150 nodes, LALARPL still maintains its lead. At a packet arrival rate of 0.1, its PDR of 0.9 is $7.06\%$ higher than the other protocols' average PDR (0.84). Even at a higher $\lambda$ of 0.2, LALARPL's PDR of 0.86 is $10.81\%$ better than the others' average PDR (0.77).

\begin{figure}
    \centering
    \includegraphics[width=0.5\linewidth]{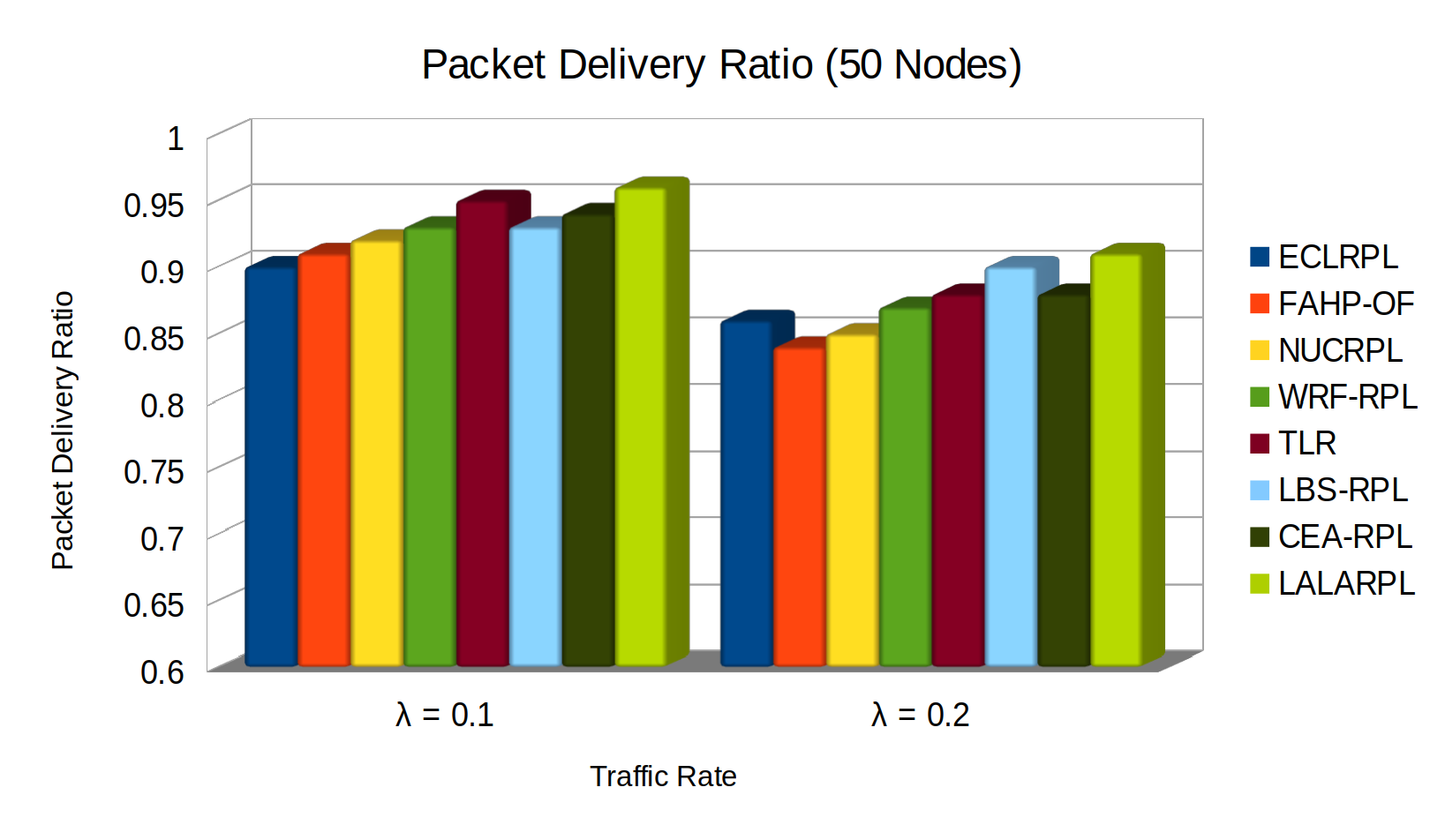}
    \caption{Packet Delivery Ratio in 150 nodes test }
    \label{fig:pdr-50}
\end{figure}

\begin{figure}
    \centering
    \includegraphics[width=0.5\linewidth]{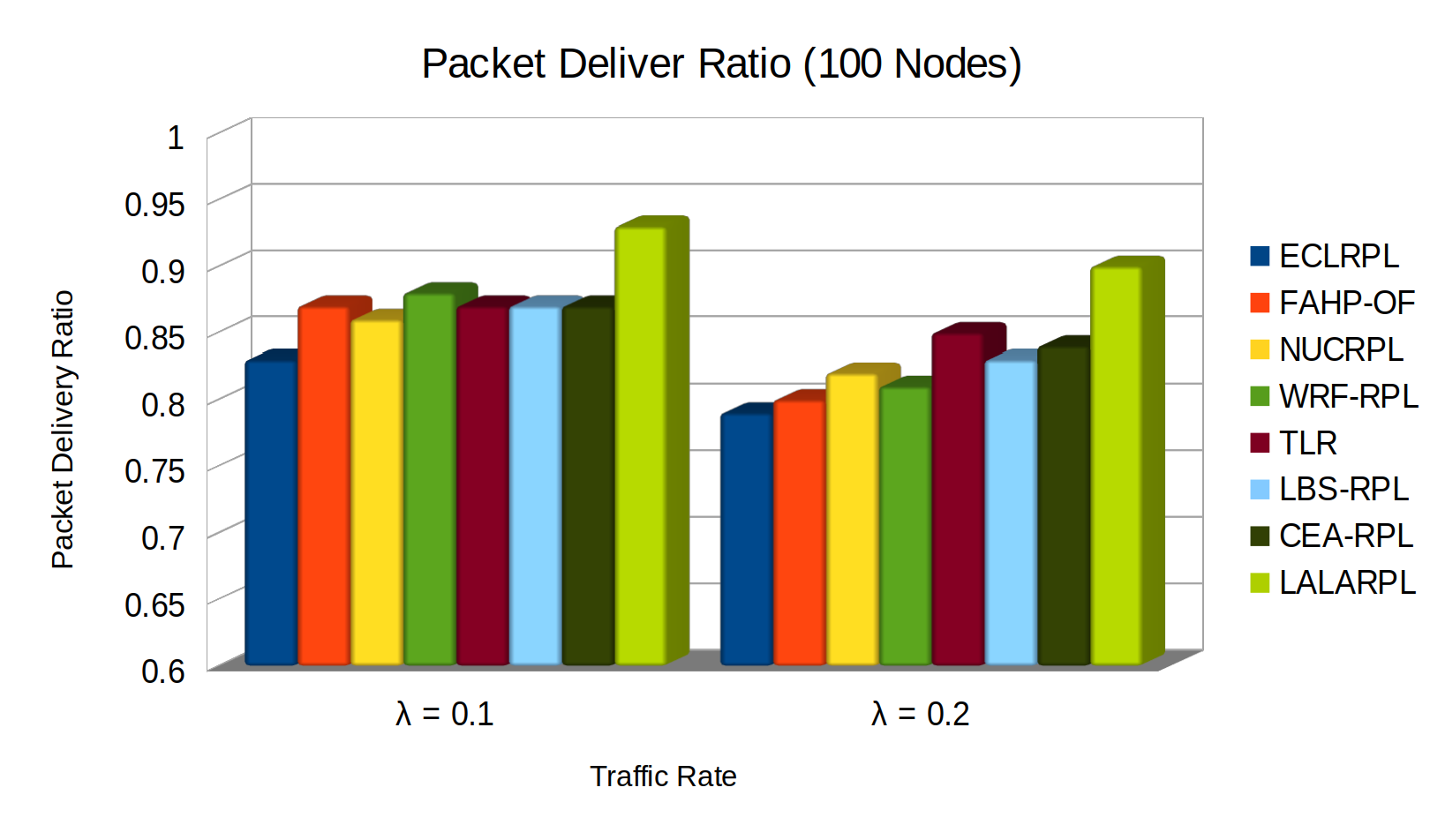}
    \caption{Packet Delivery Ratio in 100 nodes test}
    \label{fig:pdr-100}
\end{figure}

\begin{figure}
    \centering
    \includegraphics[width=0.5\linewidth]{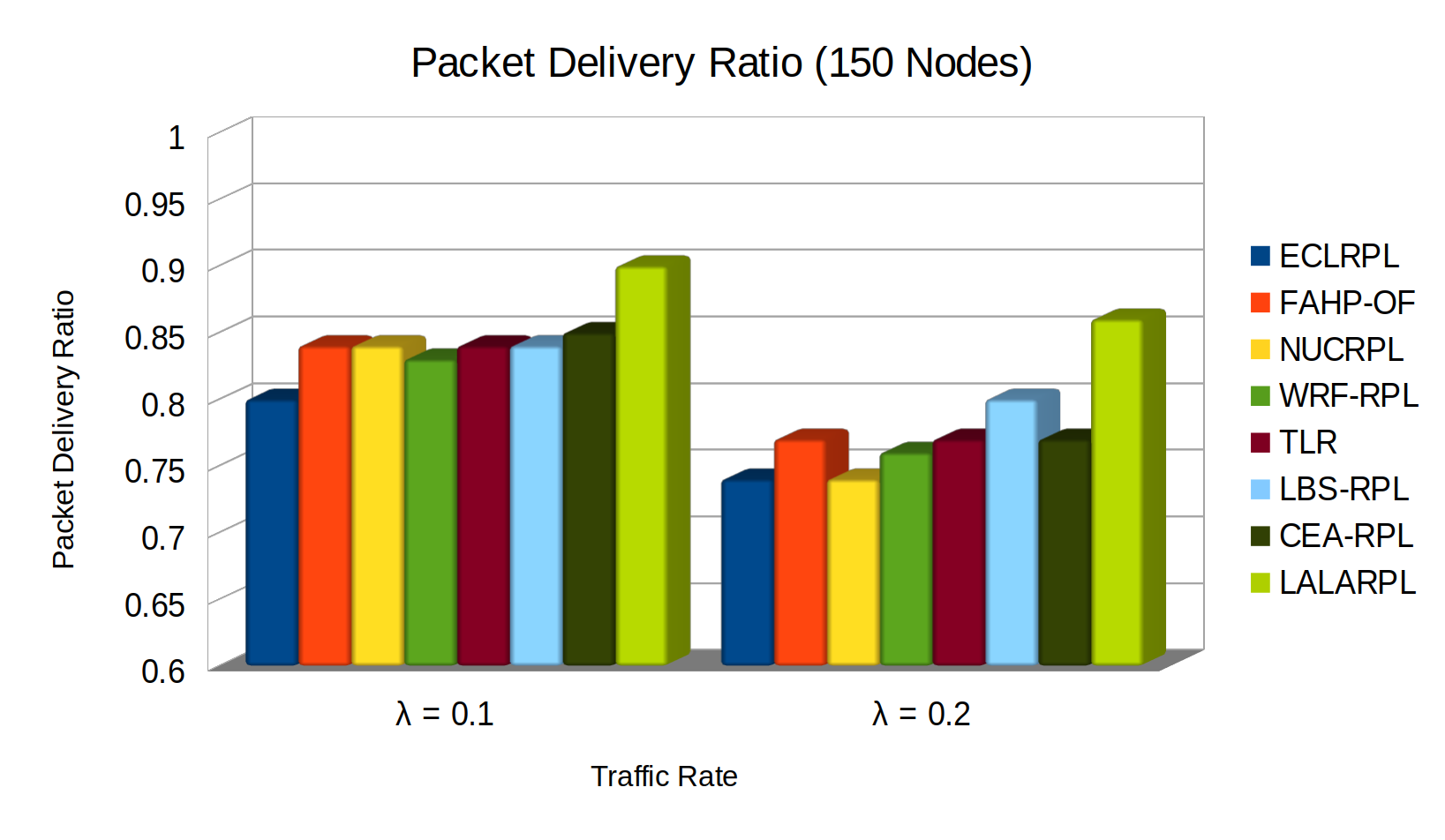}
    \caption{Packet Delivery Ratio in 150 nodes test}
    \label{fig:pdr-150}
\end{figure}

These results, depicted in Figures \ref{fig:pdr-50}, \ref{fig:pdr-100}, and \ref{fig:pdr-150} for 50, 100, and 150 node networks, respectively, underscore the robustness of LALARPL against network scaling and increased traffic. The tabulated and graphically represented data conclusively demonstrate that LALARPL improves PDR in isolation and consistently across varying network conditions, cementing its position as a formidable protocol for load balancing in RPL-based IoT networks.

\clearpage


\clearpage

\subsection{Throughput}
To derive the throughput for each node, one must consider the node's data reception and transmission over a specific period. We define the throughput, denoted as \( \theta \), for the \( i \)-th node in an RPL network. The throughput can be affected by network parameters such as link quality, number of child nodes, and the traffic pattern. Here's the expression for throughput (Equation \ref{eq:throughput}):

\begin{equation}
\theta_i = \frac{\sum R_{ij}}{\Delta t}
\label{eq:throughput}
\end{equation}

where \( \theta_i \) is the throughput for the \( i \)-th node, \( R_{ij} \) represents the data received from node \( j \) during the time interval \( \Delta t \).

Where:
\begin{itemize}
  \item \( \theta_i \) is the throughput of the \( i \)-th node.
  \item \( R_{ij} \) represents the total packets received by node \( i \) from its \( j \)-th neighbour.
  \item \( \Delta t \) is the time interval the throughput is measured.
\end{itemize}

Formula ~\ref{eq:throughput} captures the packets a node receives from all its neighbours (children) over a given period, a direct measure of throughput. This essential measurement can include other factors like packet size if needed.
In your RPL network considerations for Load Balancing, load balancing aims to evenly distribute traffic among all available nodes to prevent any single node from becoming a bottleneck, which would decrease overall network performance. 
Considering factors such as link quality and the number of children of a parent node, we update the throughput formula for node \(i\) in an RPL network and present it as follows:

\begin{equation}
\theta_i = \frac{\sum (R_{ij} \times LQI_{ij})}{\Delta t} \times \log(1 + C_i)
\label{eq:advanced_throughput}
\end{equation}

Where:
\begin{itemize}
  \item \( LQI_{ij} \) is the link quality indicator between node \( i \) and its \( j \)-th neighbour, which affects the reliability and speed of the transmitted data.
  \item \( C_i \) is the number of child nodes for node \( i \), influencing the load handled by the node.
\end{itemize}

Equation~\ref{eq:advanced_throughput} considers the quantity of data transmitted, the quality of the links, and the structural load on the node, providing a more comprehensive measure of throughput in a load-balanced RPL network. Throughput is a critical metric in assessing the performance of network protocols, particularly in IoT environments where data efficiency is paramount. The LALARPL protocol has demonstrated notable advancements in this regard, as evidenced by its superior throughput rates in simulations with 50, 100, and 150 nodes. By imposing a cap on the number of parent nodes in the children’s list, LALARPL reduces the overhead involved in parent selection and streamlines the decision-making process. This constraint simplifies the network topology and minimizes the chance of creating suboptimal paths, thus enhancing the overall data transmission efficiency.

LALARPL's dynamic weight assignment for parent nodes, guided by learning automata that adapt based on traffic and congestion, also allows for a responsive and flexible adaptation to changing network conditions. This feature ensures that traffic is distributed more evenly across the network, reducing bottlenecks and enhancing throughput.

In quantitative terms, LALARPL exhibits a remarkable improvement in throughput percentages across different network scales and packet intervals. At 50 nodes with $\lambda = 0.1$, LALARPL's throughput is 4.74, which is $7.36\%$ higher than the average throughput of 4.41 from other protocols. With a higher packet interval ($\lambda = 0.2$), the improvement is $10.35\%$ over the average throughput of 8.75. For a network of 100 nodes, the throughput under LALARPL is 10.11 at $\lambda = 0.1$, which is $3.26\%$ higher than the average of 9.79 for the others. At $\lambda = 0.2$, LALARPL achieves 19.35, presenting a robust $7.38\%$ improvement over the average throughput of 18.02. At the highest scale tested, with 150 nodes, LALARPL continues to outshine its counterparts. For $\lambda = 0.1$, its throughput of 15.94 surpasses the average of other protocols (13.29) by $19.95\%$. At $\lambda = 0.2$, the improvement is a significant $11.61\%$ over the average throughput of 26.49.

\begin{figure}
    \centering
    \includegraphics[width=0.5\linewidth]{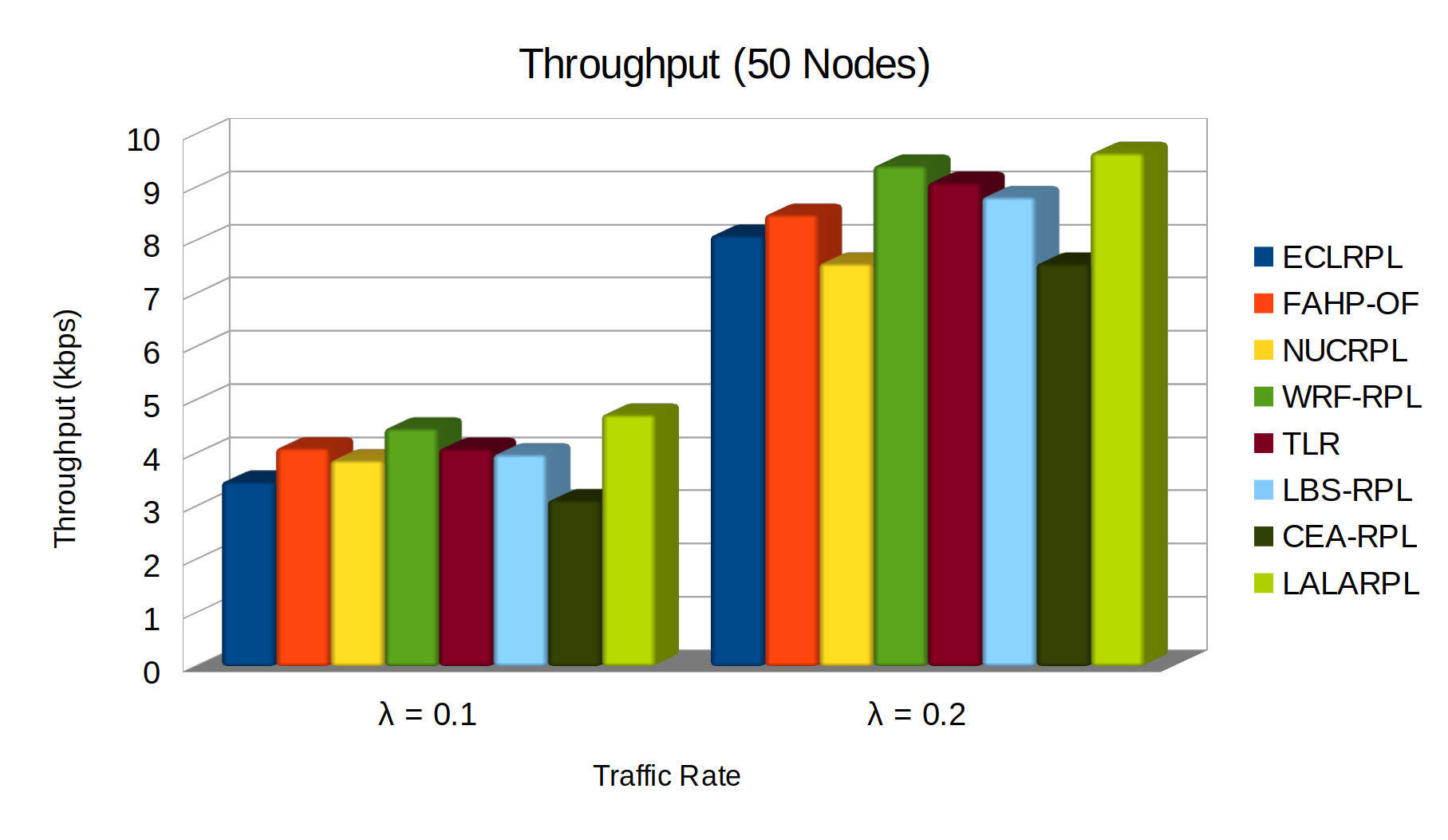}
    \caption{Throughput in 50 nodes test}
    \label{fig:throughput_50_nodes}
\end{figure}

\begin{figure}
    \centering
    \includegraphics[width=0.5\linewidth]{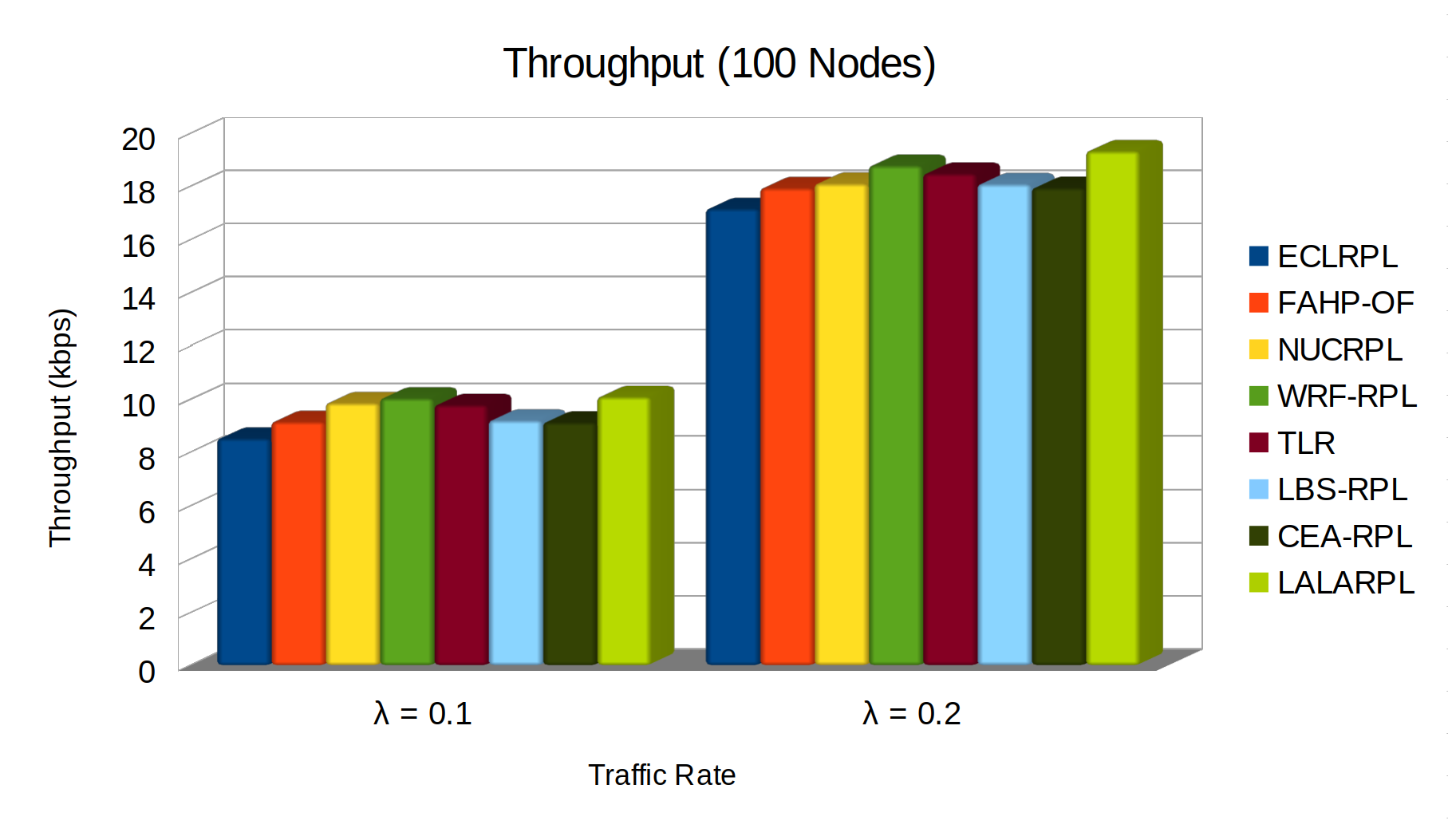}
    \caption{Throughput in 100 nodes test}
    \label{fig:throughput_100_nodes}
\end{figure}

\begin{figure}
    \centering
    \includegraphics[width=0.5\linewidth]{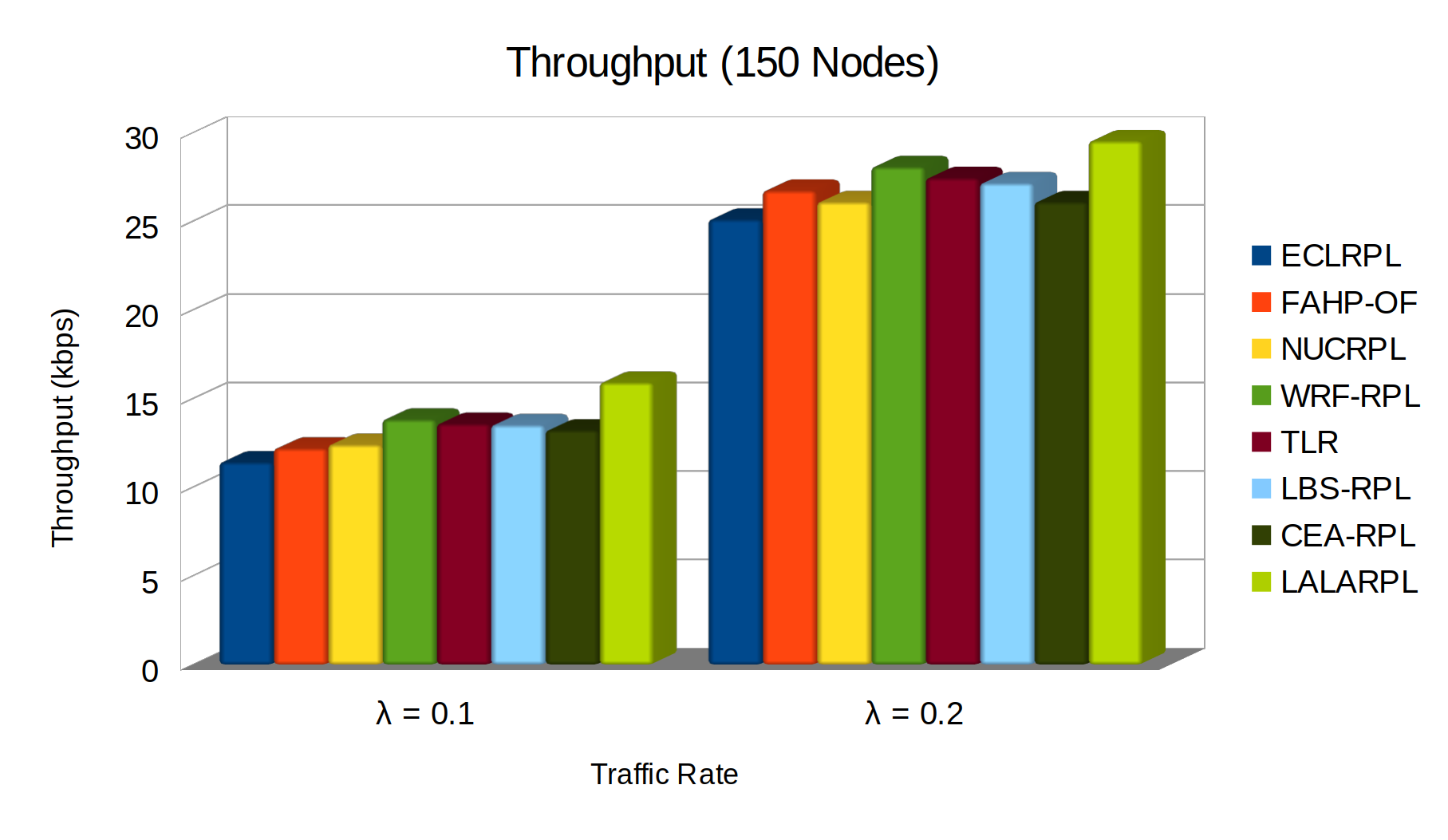}
    \caption{Throughput in 150 nodes test}
    \label{fig:throughput_150_nodes}
\end{figure}

Figures \ref{fig:throughput_50_nodes}, \ref{fig:throughput_100_nodes}, and \ref{fig:throughput_150_nodes} in the study illustrate these results in detail, showcasing the throughput outcomes for networks with 50, 100, and 150 nodes, respectively. LALARPL's consistent performance advantage underscores the protocol's effectiveness in handling network traffic and its capability to sustain higher throughput levels, solidifying its potential for large-scale IoT deployments.

\clearpage

\clearpage

\subsection{Jain Fairness Index in Throughput}

Integrating advanced throughput calculations into evaluating JFI provides a nuanced understanding of network performance, highlighting efficiency and equity in resource distribution within a load-balanced RPL context.
To quantify both the throughput of a child link \(i\) to a parent \(p\) and the fairness in resource allocation, the throughput and fairness are defined using equation~\ref{eq:advanced_throughput}. Subsequently, the Jain Fairness Index in Throughput is calculated using the following formula (Equation~\ref{eq:jain-fairness-index-throughput}):

\begin{equation}
F = \frac{\left(\sum_{i=1}^n \theta_i \right)^2}{n \cdot \sum_{i=1}^n \theta_i^2}
\label{eq:jain-fairness-index-throughput}
\end{equation}

Where:
\begin{itemize}
    \item \( \theta_i \) denotes the throughput for the \(i\)-th node, considering the data transmitted, the quality of the links (LQI), and the number of child nodes (C).
    \item The numerator \(\left(\sum_{i=1}^n \theta_i \right)^2\) is the square of the sum of the throughputs for all nodes.
    \item The denominator \(n \cdot \sum_{i=1}^n \theta_i^2\) represents the product of the number of nodes and the sum of the squares of individual throughputs, reflecting the dispersion of throughput across the network and facilitating the assessment of the load balancing effectiveness in terms of fairness.
\end{itemize}

The JFI offers an invaluable metric for assessing the equity of resource allocation among network nodes, particularly regarding throughput. A higher JFI indicates a more equitable distribution of network capacity among nodes, which is crucial for the consistent performance of applications in an IoT environment. The LALARPL protocol makes notable strides in the fairness of throughput distribution. By limiting each child node to a maximum of five parent nodes, the protocol ensures that node throughput remains high without overburdening any single node, thus promoting fair resource allocation and mitigating potential bottlenecks. Incorporating learning automata within LALARPL enhances the protocol's adaptive capabilities, allowing for dynamic adjustments based on real-time network conditions. This adaptability is especially beneficial for managing network traffic, avoiding congestion, and preventing bottlenecks—all of which contribute to maintaining high JFI values.

By examining the provided simulation data, LALARPL demonstrates superior fairness in various network configurations and packet intervals. For a network with 50 nodes at $\lambda = 0.1$ packet/s, LALARPL achieves a JFI of 0.89, which is $2.56\%$ higher than the average JFI of 0.868 of other protocols. The improvement is substantial at $\lambda = 0.2$ packet/s, at $5.88\%$ over the average JFI of 0.822. When the network scales up to 100 nodes, the JFI improvement is still evident. At $\lambda = 0.1$ packet/s, LALARPL's JFI of 0.93 is $4.49\%$ higher than the average JFI of 0.89 of the other protocols. At the increased packet interval ($\lambda = 0.2$), LALARPL's JFI of 0.91 represents a considerable $9.64\%$ improvement over the average JFI of 0.83. At 150 nodes, LALARPL's advantages persist. With $\lambda = 0.1$ packet/s, LALARPL's JFI of 0.88 is $4.76\%$ better than the average JFI of 0.839. At a packet interval of $\lambda = 0.2$ packet/s, LALARPL records a JFI of 0.87, showcasing an impressive $13.33\%$ improvement over the average JFI of 0.768.

\begin{figure}
    \centering
    \includegraphics[width=0.5\linewidth]{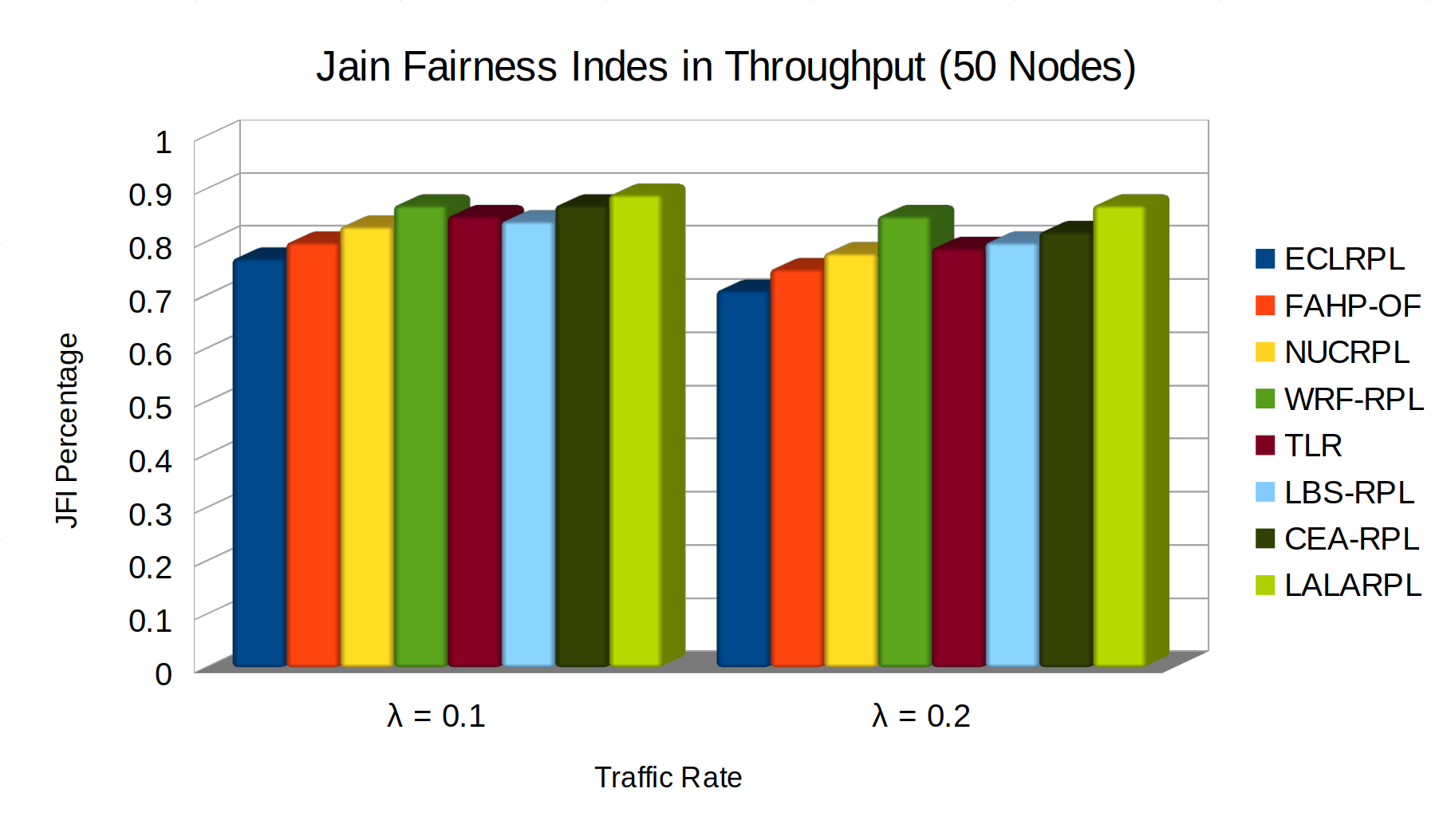}
    \caption{Jain Fairness Index of Throughput in 50 nodes test}
    \label{fig:JFI-throughput_50_nodes}
\end{figure}

\begin{figure}
    \centering
    \includegraphics[width=0.5\linewidth]{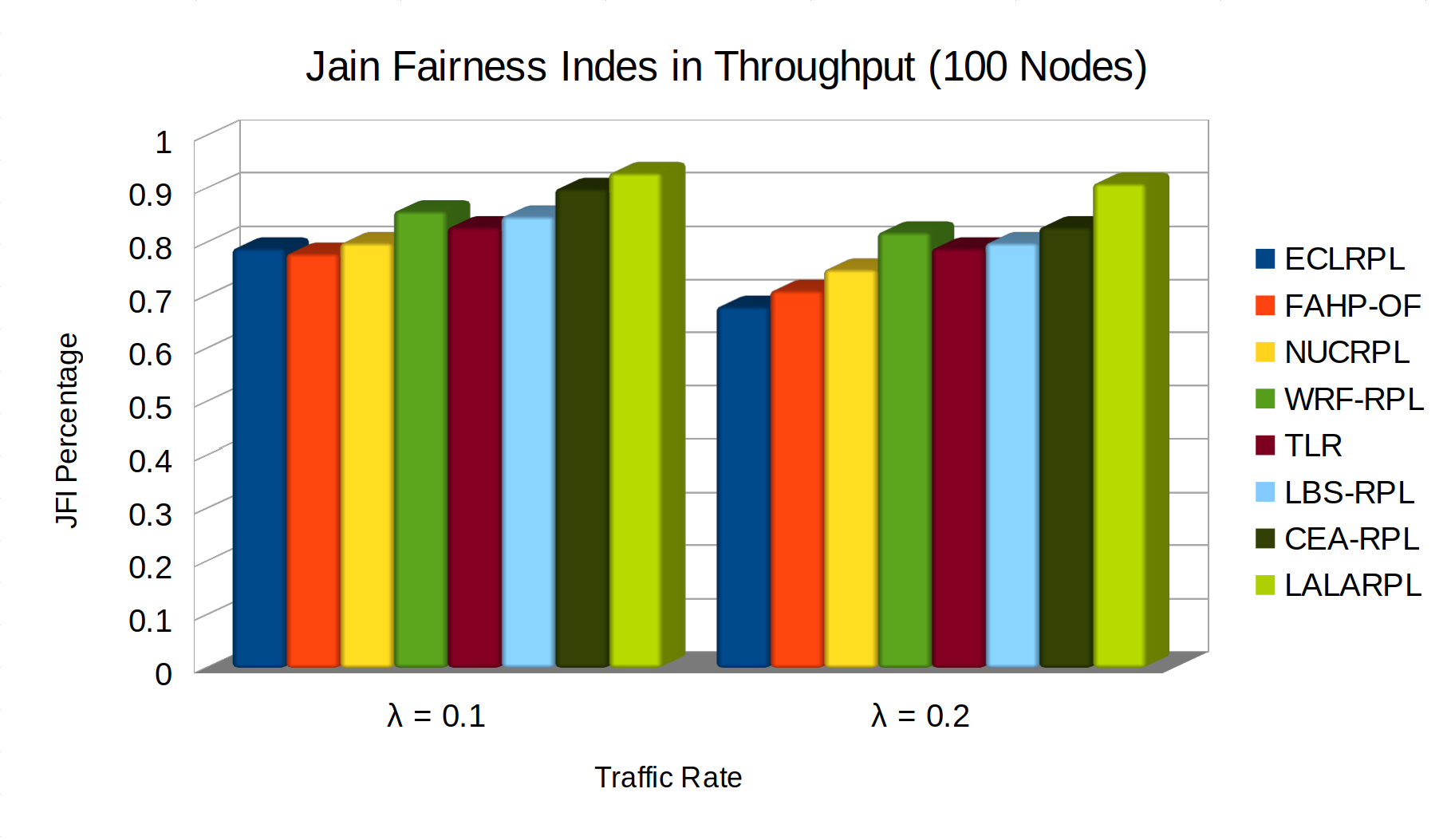}
    \caption{Jain Fairness Index of Throughput in 100 nodes test}
    \label{fig:JFI-throughput_100_nodes}
\end{figure}

\begin{figure}
    \centering
    \includegraphics[width=0.5\linewidth]{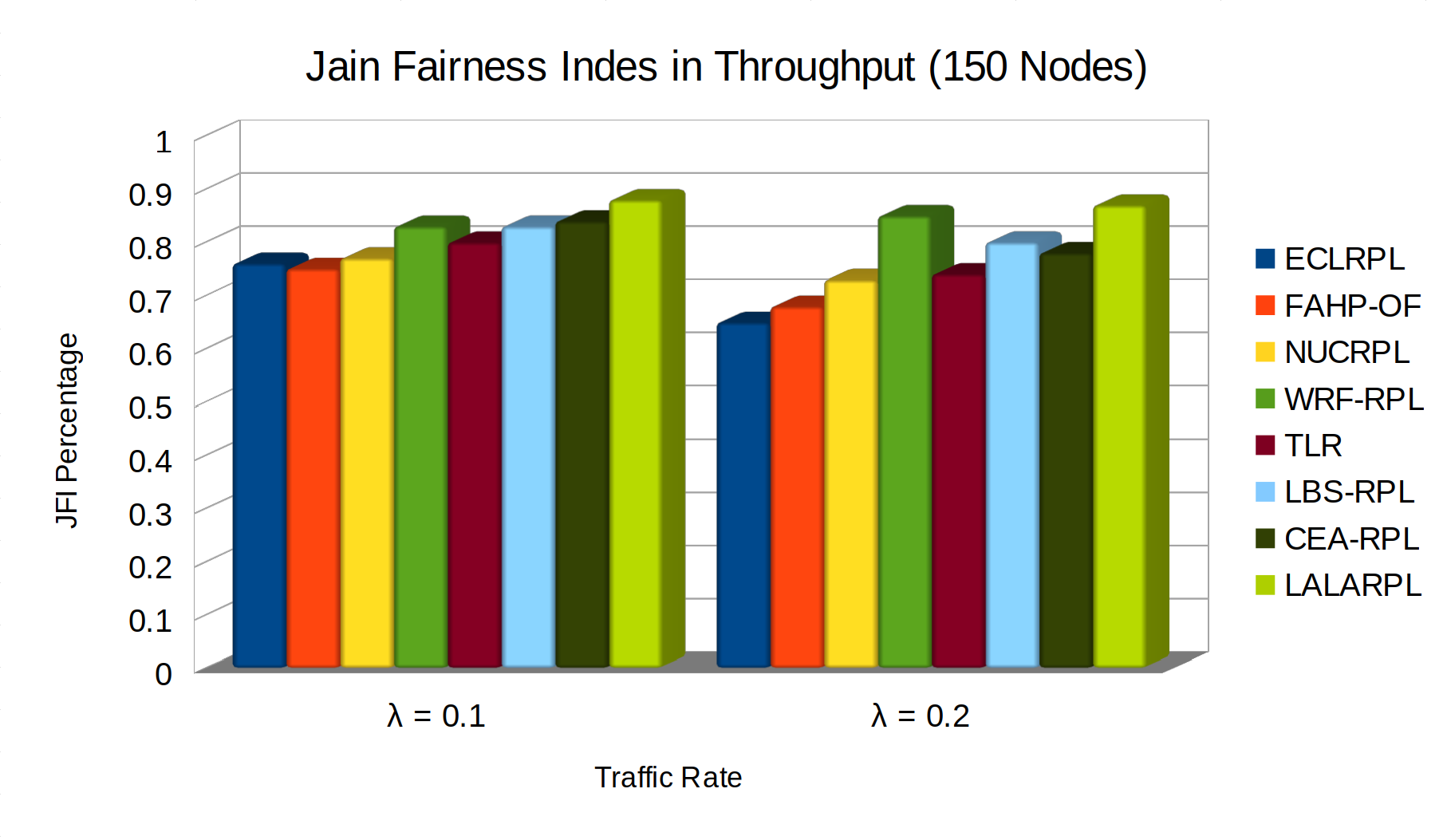}
    \caption{Jain Fairness Index of Throughput in 150 nodes test}
    \label{fig:JFI-throughput_150_nodes}
\end{figure}

These fairness improvements demonstrate the effectiveness of LALARPL in managing network resources and indicate its potential for reducing the negative impacts of increased traffic load and congestion. By employing a learning-based approach, LALARPL can continuously optimize its behaviour to sustain fairness and efficiency, even as network dynamics evolve. The results of these simulations are visually detailed in Figures \ref{fig:JFI-throughput_50_nodes}, \ref{fig:JFI-throughput_100_nodes}, and \ref{fig:JFI-throughput_150_nodes} for networks of 50, 100, and 150 nodes, respectively, reflecting the robust and equitable throughput performance of LALARPL across different network scales and traffic conditions.

\clearpage

\clearpage

\subsection{Average End to End Delay}

The Average End-to-End Delay (AEED) is a vital statistic that reflects the time data packets travel from the source to the destination in a network. It encapsulates the propagation and transmission delays and the processing time at each hop along the route, queuing delays, and possible retransmissions due to errors or packet losses.
Latency in networking is defined as the time required for a packet to traverse from the source to its destination. Unlike some methodologies that compute this metric in an aggregated end-to-end manner, our proposed method analyzes latency step-by-step. This approach ensures that the latency of a packet encompasses the cumulative delays incurred at each node, including processing, queuing, transmission, and propagation times. As a result, minimizing these individual components effectively reduces the overall network latency. The delay at each node \( i \) is detailed in the following equation ~\ref{eq:node-delay}:
\begin{equation}
{NodeDelay}_i = {Proc\_Delay}_i + {Queue\_Delay}_i + {Trans\_Delay}_i + {Prop\_Delay}_i
\label{eq:node-delay}
\end{equation}

Where:
\begin{itemize}
  \item \( {Proc\_Delay}_i \) is the processing delay at node \( i \),
  \item \( {Queue\_Delay}_i \) is the queuing delay at node \( i \),
  \item \( {Trans\_Delay}_i \) is the transmission delay at node \( i \),
  \item \( {Prop\_Delay}_i \) is the propagation delay from node \( i \).
\end{itemize}

Additionally, the latency experienced over each link for parent \( p \) is represented through the Link Delay Index, which aggregates the delays across different links to provide a metric for assessing packet transmission efficiency, especially in dynamic networking environments. This index is defined as follows (Equation ~\ref{eq:link-delay}):
\begin{equation}
LinkDelayIndex_i^p = \sum {Delay_i^p}
\label{eq:link-delay}
\end{equation}

Another critical metric in IoT networks that utilize the RPL protocol is the average end-to-end delay, which assesses the network's responsiveness and operational efficiency. This delay, denoted as \( \Delta \), is computed using the formula (Equation ~\ref{eq:avg-end2end-delay}):
\begin{equation}
\Delta = \frac{1}{N} \sum_{i=1}^N \tau_i
\label{eq:avg-end2end-delay}
\end{equation}

Where:
\begin{itemize}
  \item \( N \) is the total number of packets successfully delivered during the observation period,
  \item \( \tau_i \) represents the time the \( i \)-th packet travels from its source to its destination.
\end{itemize}

LALARPL's commendable performance in reducing AEED can be attributed to several innovative protocol aspects. By limiting the number of potential parent nodes for any child node, LALARPL ensures a less congested and more streamlined path selection process, thereby reducing queuing delays. Furthermore, the protocol's learning automata dynamically adjust the weights of various routing parameters, such as link quality, node congestion levels, and transmission rates, which results in a more efficient route selection and fewer packet retransmissions.

Analyzing the simulation results, we can deduce that LALARPL consistently outperforms the other protocols in terms of AEED across different network sizes and packet intervals. Specifically, in a network with 50 nodes at $\lambda = 0.1$ packet/s, LALARPL achieves an AEED of 11.064 milliseconds, which is approximately $5.14\%$ lower than the following best protocol, WRF-RPL, with an AEED of 11.089 milliseconds. At a packet interval of $\lambda = 0.2$ packet/s, the improvement is more pronounced at $11.88\%$ over the average AEED of 17.33 milliseconds of the other protocols. In a larger network of 100 nodes, LALARPL continues to demonstrate its efficacy. At $\lambda = 0.1$ packet/s, LALARPL's AEED is 13.564 milliseconds, which is $7.52\%$ better than the average AEED of 14.662 milliseconds of the competing protocols. For $\lambda = 0.2$ packet/s, LALARPL's performance improvement is $10.06\%$ over the average AEED of 23.761 milliseconds. Scaling up to 150 nodes, the advantage of LALARPL is still significant. With $\lambda = 0.1$ packet/s, LALARPL's AEED is 16.061 milliseconds, offering a $4.92\%$ improvement over the average AEED of 16.887 milliseconds of the other protocols. At $\lambda = 0.2$ packet/s, LALARPL presents a $7.07\%$ better AEED, measuring 26.245 milliseconds compared to the average of 28.235 milliseconds.

\begin{figure}
    \centering
    \includegraphics[width=0.5\linewidth]{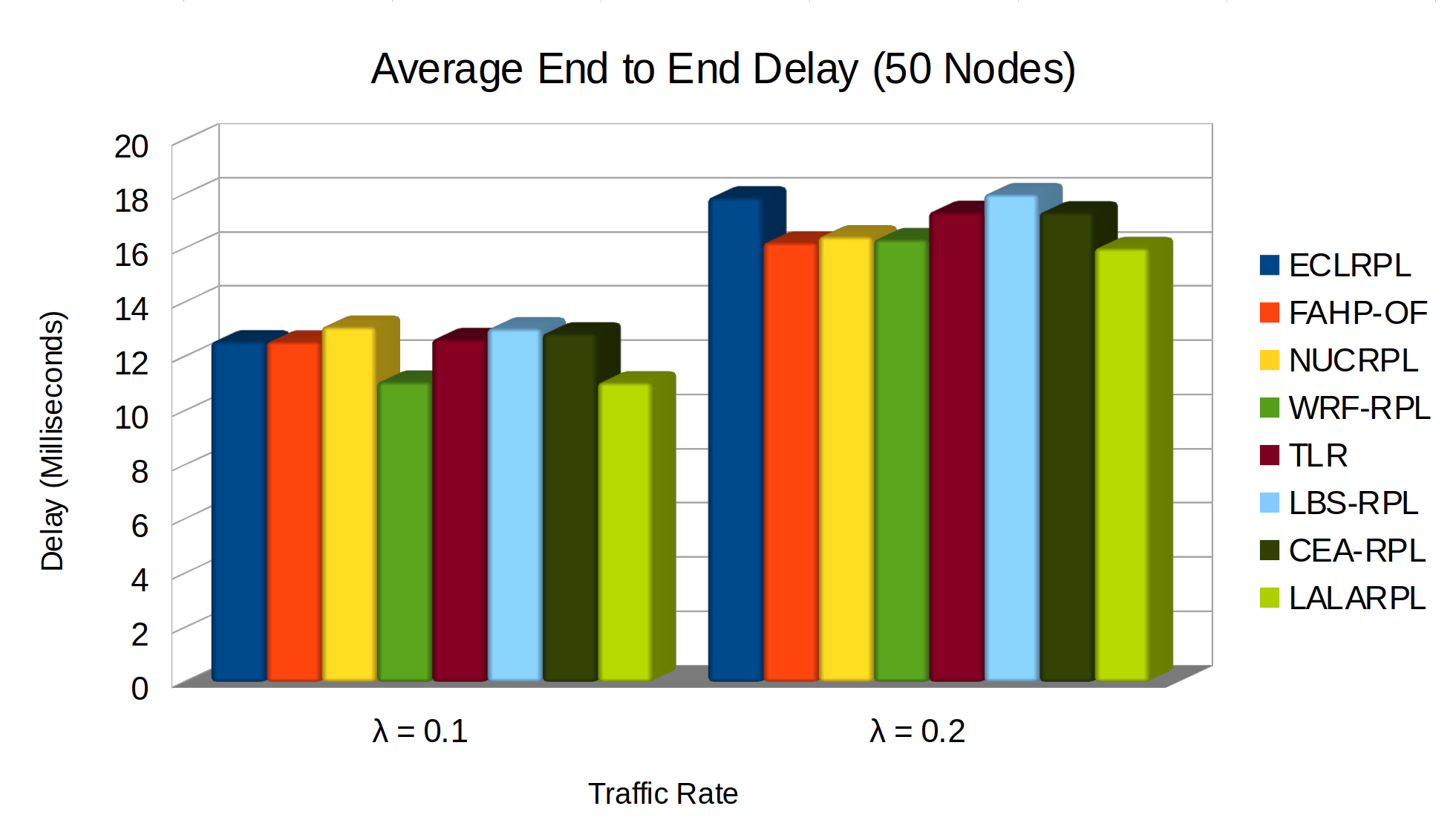}
    \caption{Average End to End delay in 50 nodes test}
    \label{fig:End2end-delay_50_nodes}
\end{figure}

\begin{figure}
    \centering
    \includegraphics[width=0.5\linewidth]{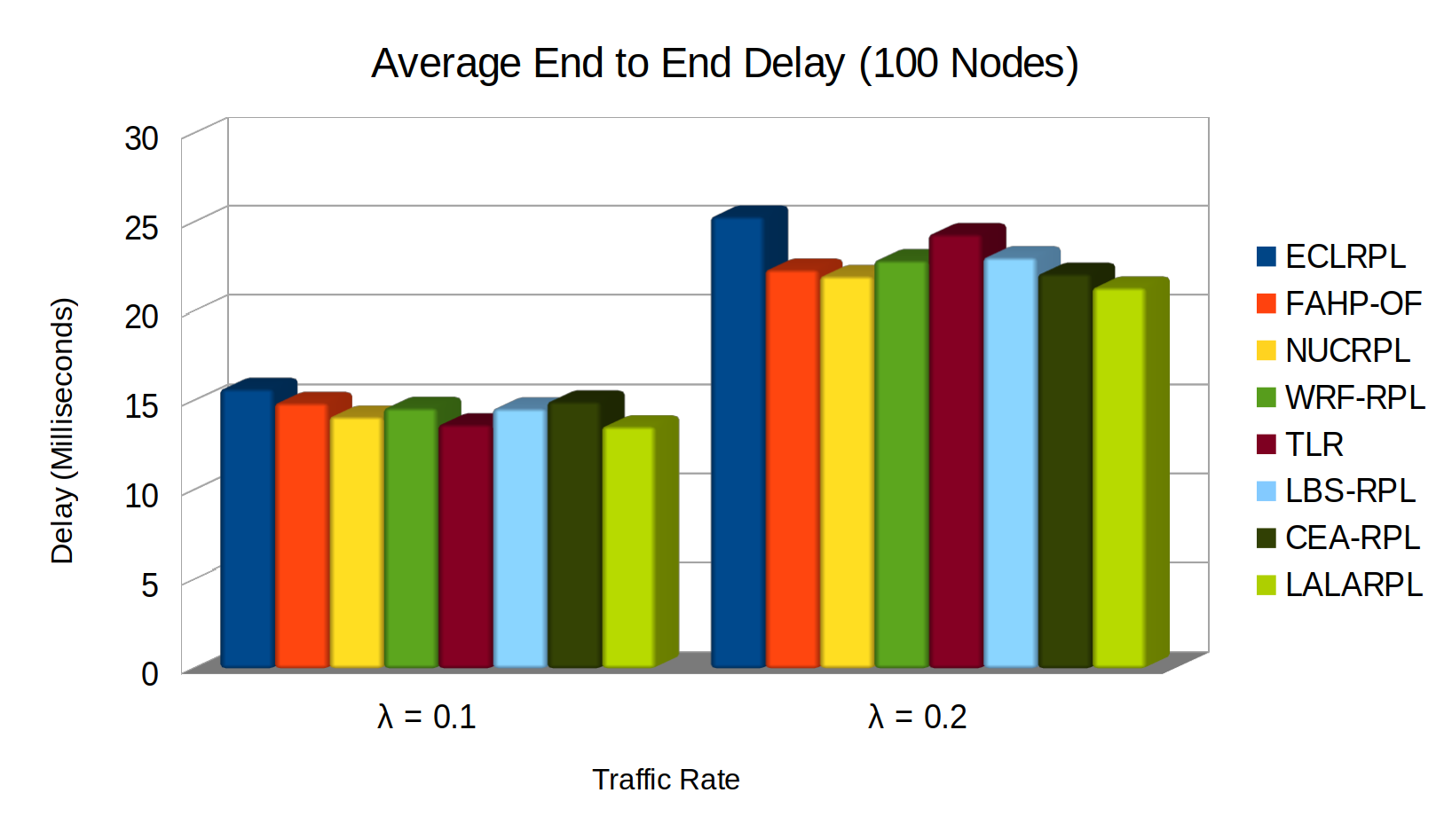}
    \caption{Average End to End delay in 100 nodes test}
    \label{fig:End2end-delay_100_nodes}
\end{figure}

\begin{figure}
    \centering
    \includegraphics[width=0.5\linewidth]{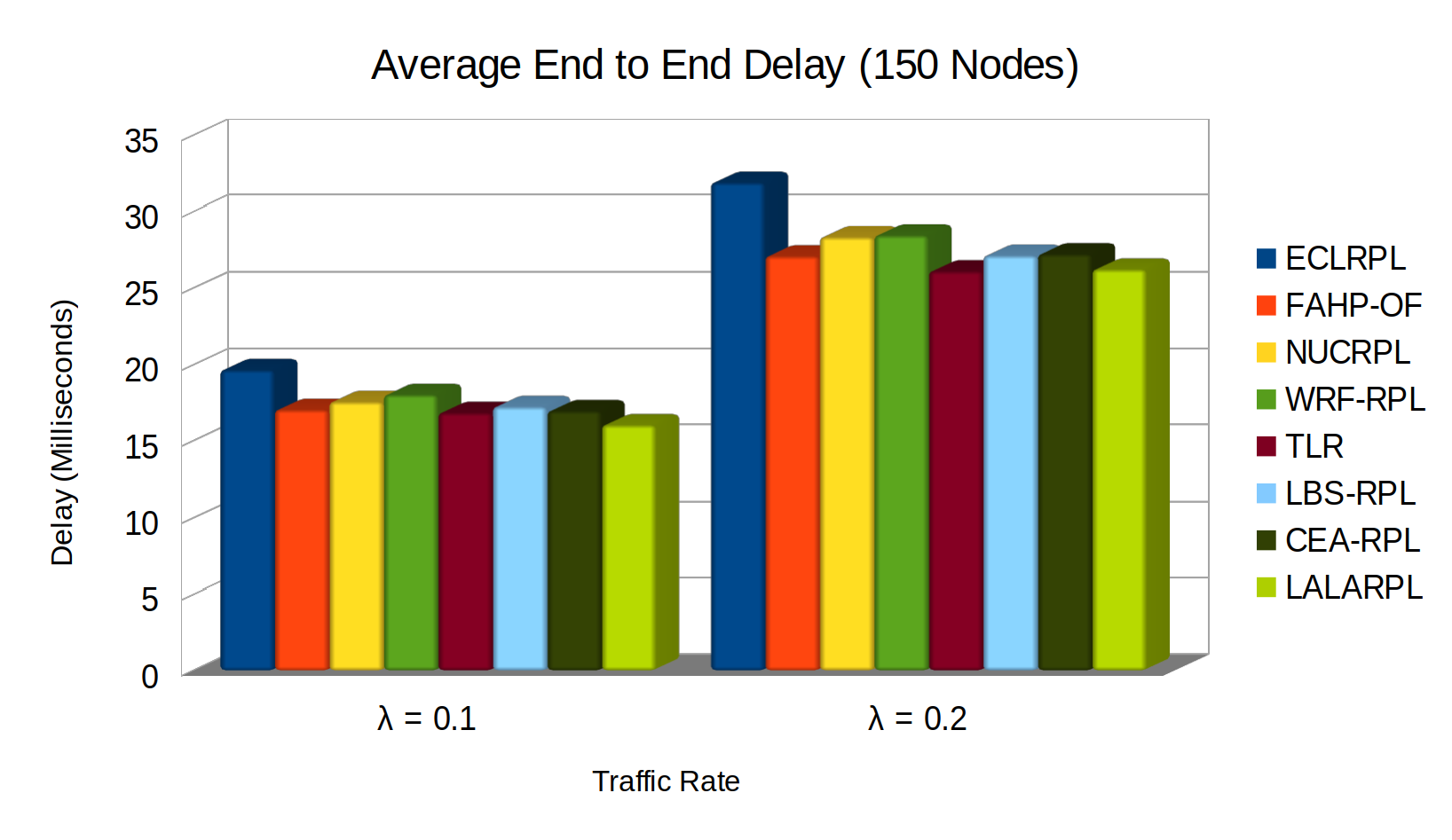}
    \caption{Average End to End delay in 150 nodes test}
    \label{fig:End2end-delay_150_nodes}
\end{figure}

The cumulative effect of these optimizations is clearly illustrated in Figures \ref{fig:End2end-delay_50_nodes}, \ref{fig:End2end-delay_100_nodes}, and \ref{fig:End2end-delay_150_nodes} (referencing the figures provided for AEED results in networks of 50, 100, and 150 nodes, respectively). These improvements are not only indicative of LALARPL’s potential in enhancing network timeliness but also highlight its robustness in reducing latency, particularly in IoT environments that demand timely data delivery.

\clearpage

\clearpage

\subsection{JFI in Energy Consumption}
Given the limited power resources in such environments, energy consumption in RPL networks for IoT applications is critical. This metric is influenced by the energy expended during transmission, reception, idle, and sleep modes. The total energy consumption \( E_{\text{total}} \) for node \( i \) can be calculated by considering the power consumed in different operational states. The equation \ref{eq:Energy-total} describes this:
\begin{equation}
E_{\text{total}_i} = P_T \cdot t_T + P_R \cdot t_R + P_I \cdot t_I + P_S \cdot t_S
\label{eq:Energy-total}
\end{equation}

Where:
\begin{itemize}
    \item \( P_T \), \( P_R \), \( P_I \), \( P_S \) represent the power consumption while transmitting, receiving, idling, and sleeping, respectively,
    \item \( t_T \), \( t_R \), \( t_I \), \( t_S \) are the durations spent in each corresponding state.
\end{itemize}

This equation provides a comprehensive assessment of the energy utilization for each node. The Jain's Fairness Index (JFI) can be employed to evaluate the fairness of energy consumption across the network. The JFI for energy consumption \( JFI_{E} \) is calculated as follows (Equation ~\ref{eq:jain-fairness-index-energy}):
\begin{equation}
JFI_{E} = \frac{(\sum_{i=1}^n E_{\text{total}_i})^2}{n \cdot \sum_{i=1}^n E_{\text{total}_i}^2}
\label{eq:jain-fairness-index-energy}
\end{equation}

where \( n \) is the number of nodes in the network, and \( E_{total_i} \) is the total energy consumption of the \(i\)-th node.

The JFI in energy consumption provides a crucial measure of how uniformly energy resources are utilized across a network. A high JFI value signifies equitable energy usage among the nodes, which is particularly advantageous in IoT networks, where devices are often constrained by battery life. The strategic energy allocation prolongs individual nodes' operational longevity and ensures the network infrastructure's overall sustainability. The LALARPL protocol exhibits exemplary energy fairness. Its JFI scores across simulations with 50, 100, and 150 nodes at 0.1 and 0.2 packets per second packet intervals. The protocol's algorithm optimizes energy use through a combination of innovative routing decisions and dynamic load balancing:

\begin{itemize}
    \item Optimized Path Selection: LALARPL carefully chooses routes that minimize energy expenditure, thereby preserving node battery life and reducing the need for frequent transmissions.
    \item Load Distribution: By restricting the number of potential parents for a child node, LALARPL prevents the excessive energy drain of any single node, promoting a more uniform energy consumption across the network.
    \item Adaptive Learning: Incorporating learning automata in LALARPL enables the network to adapt to changing conditions and optimize the routing dynamically. This adaptation means the network can avoid energy-intensive routes that might lead to retransmissions or increased processing.
\end{itemize}

Evaluating the simulation outcomes for a network of 50 nodes, LALARPL presents a JFI of 0.92 at $\lambda = 0.1$ packet/s, surpassing the next closest protocol, WRF-RPL, by a marginal but meaningful 0.54\%. When the packet interval increases to $\lambda = 0.2$ packet/s, LALARPL maintains its lead with a JFI of 0.908, outperforming WRF-RPL's 0.91. The protocol's advantage becomes more distinct for networks of 100 nodes. LALARPL achieves a JFI of 0.965 at $\lambda = 0.1$, a noteworthy 5.47\% improvement over WRF-RPL's 0.915. At the higher packet rate, the JFI for LALARPL is 0.942, representing a 5.25\% increase over WRF-RPL. The trend of enhanced energy fairness with LALARPL continues in the 150-node network scenario. With a $\lambda$ of 0.1, LALARPL attains a JFI of 0.938, edging out WRF-RPL by 3.63\%. At $\lambda = 0.2$, LALARPL's JFI of 0.915 showcases a remarkable 4.93\% improvement over WRF-RPL.

\begin{figure}
    \centering
    \includegraphics[width=0.5\linewidth]{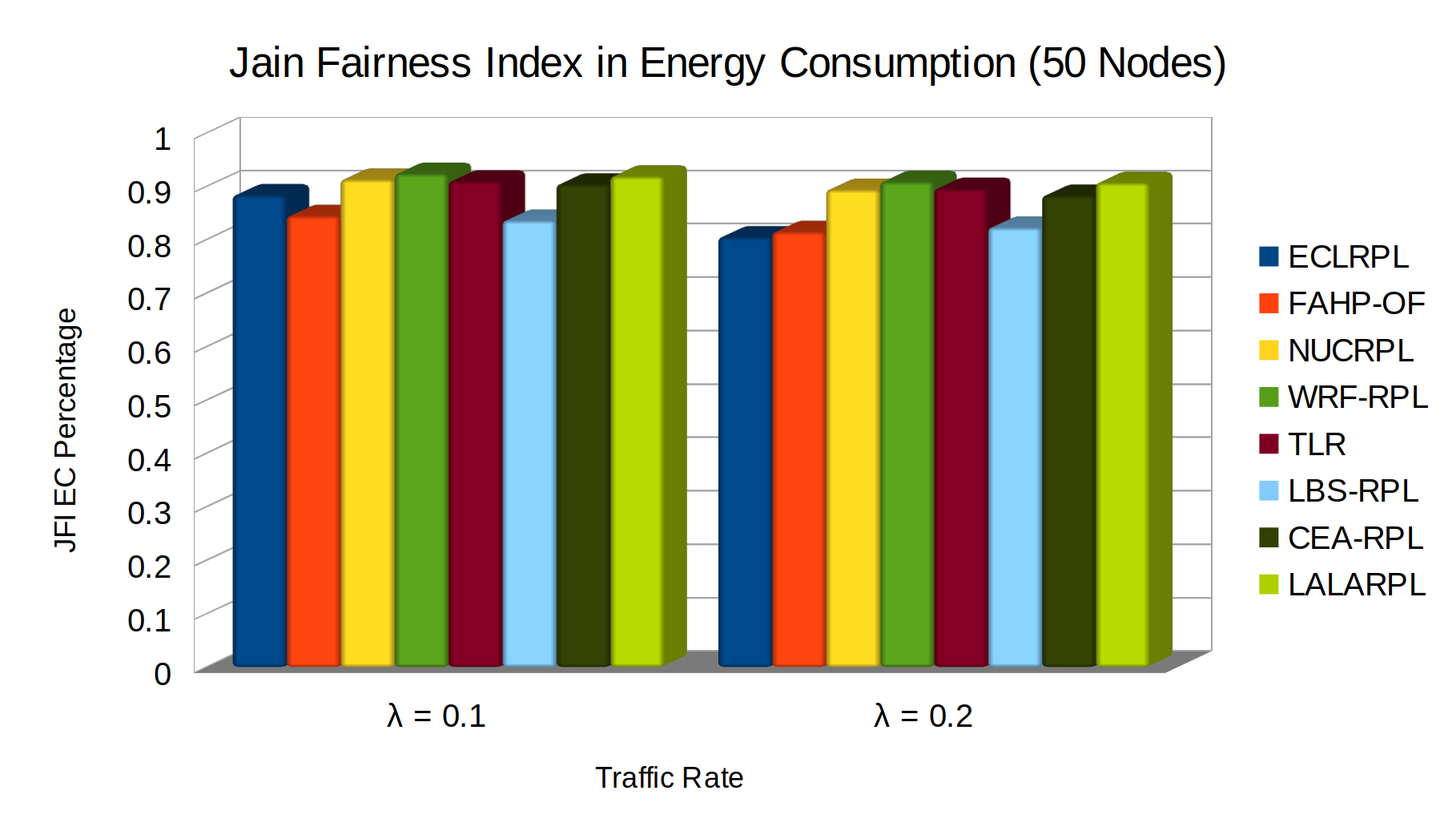}
    \caption{Jain Fairness Index of Energy in 50 nodes test}
    \label{fig:JFI-energy-50}
\end{figure}

\begin{figure}
    \centering
    \includegraphics[width=0.5\linewidth]{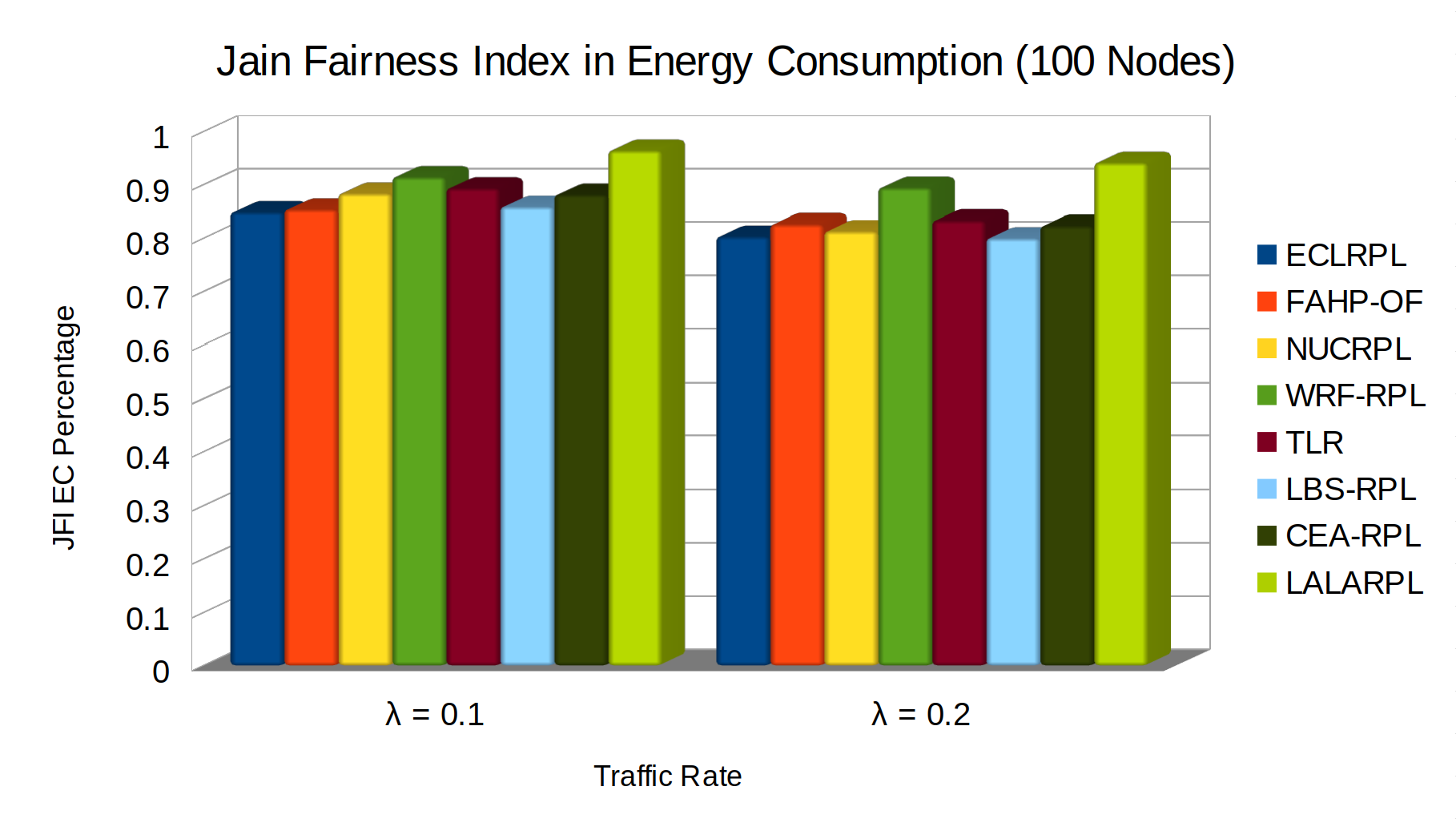}
    \caption{Jain Fairness Index of Energy in 100 nodes test}
    \label{fig:JFI-energy-100}
\end{figure}

\begin{figure}
    \centering
    \includegraphics[width=0.5\linewidth]{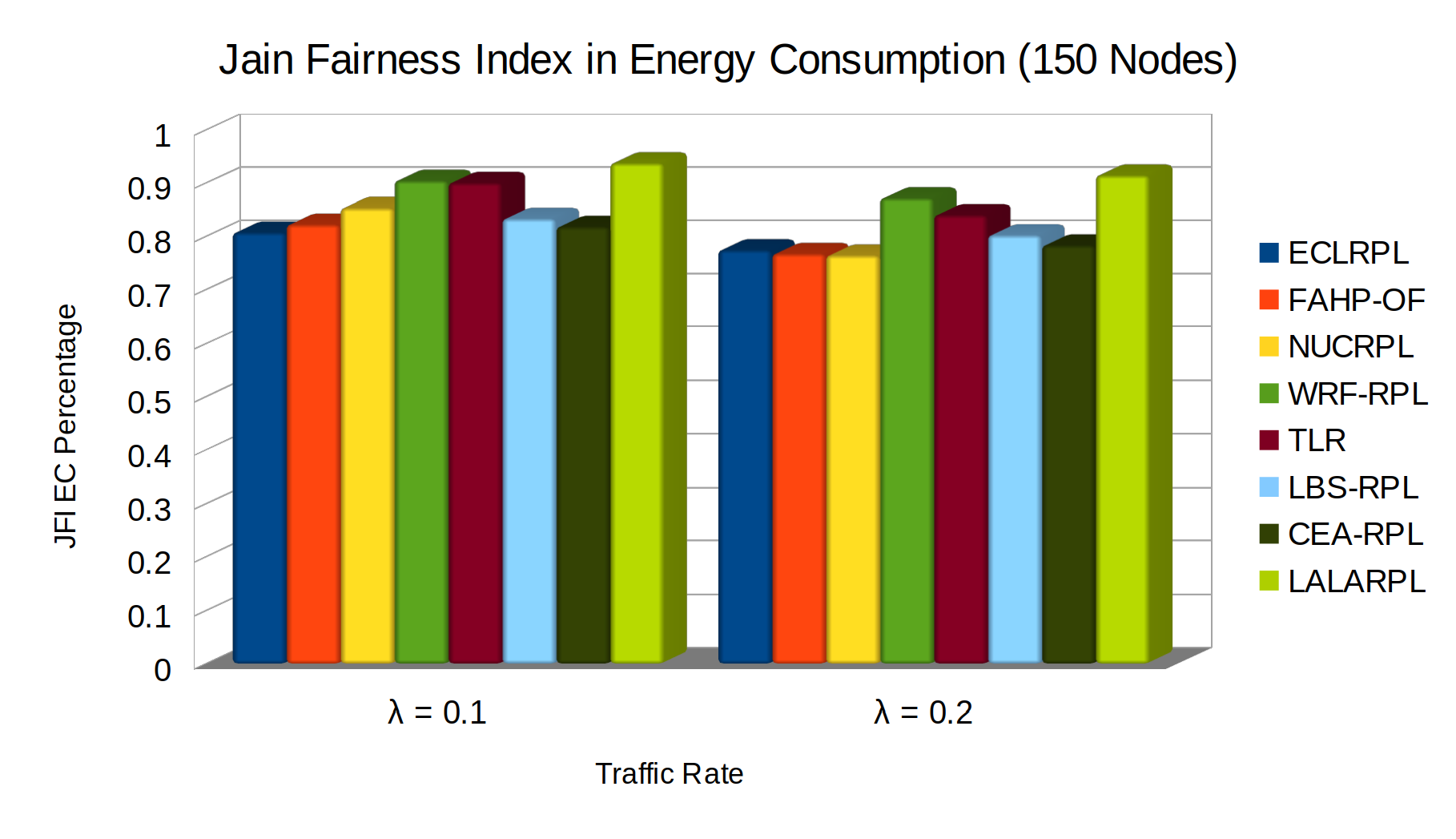}
    \caption{Jain Fairness Index of Energy in 150 nodes test}
    \label{fig:JFI-energy-150}
\end{figure}

Figures \ref{fig:JFI-energy-50}, \ref{fig:JFI-energy-100}, and \ref{fig:JFI-energy-150} visualize these enhancements in energy fairness, corresponding to the 50, 100, and 150-node networks, respectively. The data graphically represents the LALARPL protocol's superior energy distribution management, affirming its suitability for IoT applications where energy efficiency is paramount. The computed percentage improvements are based on LALARPL's highest JFI scores compared to the best-performing competing protocols in each scenario, highlighting its effectiveness in achieving energy fairness across diverse network scales and traffic conditions.

\clearpage

\clearpage

\subsection{Average Lifetime Network}
The Average Lifetime Network (ALTN)is a pivotal metric in evaluating the sustainability and operational viability of the RPL in IoT applications. This measure primarily assesses how energy resources are distributed and utilized within the network, reflecting on its ability to handle energy consumption efficiently and equitably among its nodes. The ALTN is instrumental in highlighting the effectiveness of energy management strategies within the network. By tracking the time until the death of each node, the metric provides insights into how well the network avoids or mitigates energy consumption hotspots---areas within the network where nodes may deplete their energy resources prematurely due to high traffic loads or poor distribution of network tasks. Furthermore, the delay in the time of death of the network's first node is an essential indicator of the network's overall health. A longer delay suggests that the network's design and operational protocols effectively balance energy consumption, thereby prolonging the functional period of critical network nodes.

Additionally, the ALTN metric measures the average operational lifetime across all nodes and highlights energy consumption disparities, which are critical for network optimization. Networks with a high ALTN are generally more robust and capable of sustaining operation under varying conditions, which is essential for IoT environments where continuous operation is crucial. Furthermore, analyzing ALTN helps identify nodes or areas within the network that may require redesigns, such as enhancements in routing algorithms or energy harvesting capabilities, to ensure a more balanced energy distribution and longer network lifespan. The ALTN is computed after a specific operational period, typically 1000 seconds, to evaluate the network under a standardized condition. The formula for calculating ALTN, provided in Equation~\eqref{eq:ALTN}, incorporates both the lifetimes of nodes that cease operation during the simulation and a projection for those that survive \cite{Homaei2021_69}:

\begin{equation}
ALTN = \frac{\sum_{i=1}^{N-M} t_i + (M \times \wp)}{N}
\label{eq:ALTN}
\end{equation}
where:
\begin{itemize}
    \item \( t_i \) denotes the time of death of the \(i\)-th node, crucial for understanding when each node exhausts its energy reserves.
    \item \( N \) represents the total number of nodes in the network, providing the denominator for averaging the lifetimes.
    \item \( M \) is the number of nodes still operational at the end of the simulation period, indicating the network's resilience.
    \item \( \wp \) signifies the predefined or estimated maximum lifetime for the surviving nodes, offering a way to estimate the potential maximum longevity of the network's operational capability.
\end{itemize}

\begin{figure}
    \centering
    \includegraphics[width=0.5\linewidth]{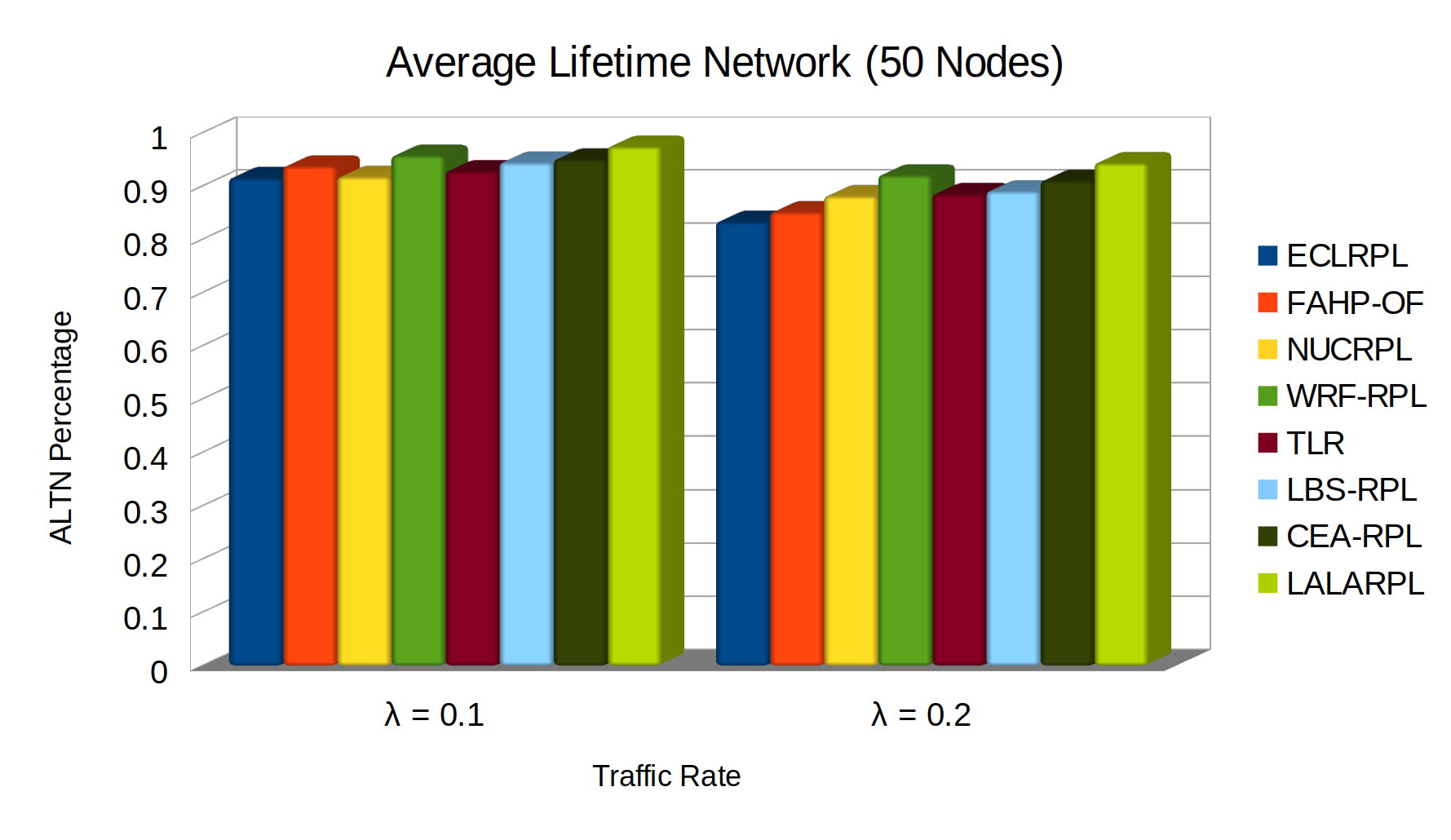}
    \caption{Average Lifetime Network in 50 nodes test}
    \label{fig:Average-Lifetime-Network-50}
\end{figure}

\begin{figure}
    \centering
    \includegraphics[width=0.5\linewidth]{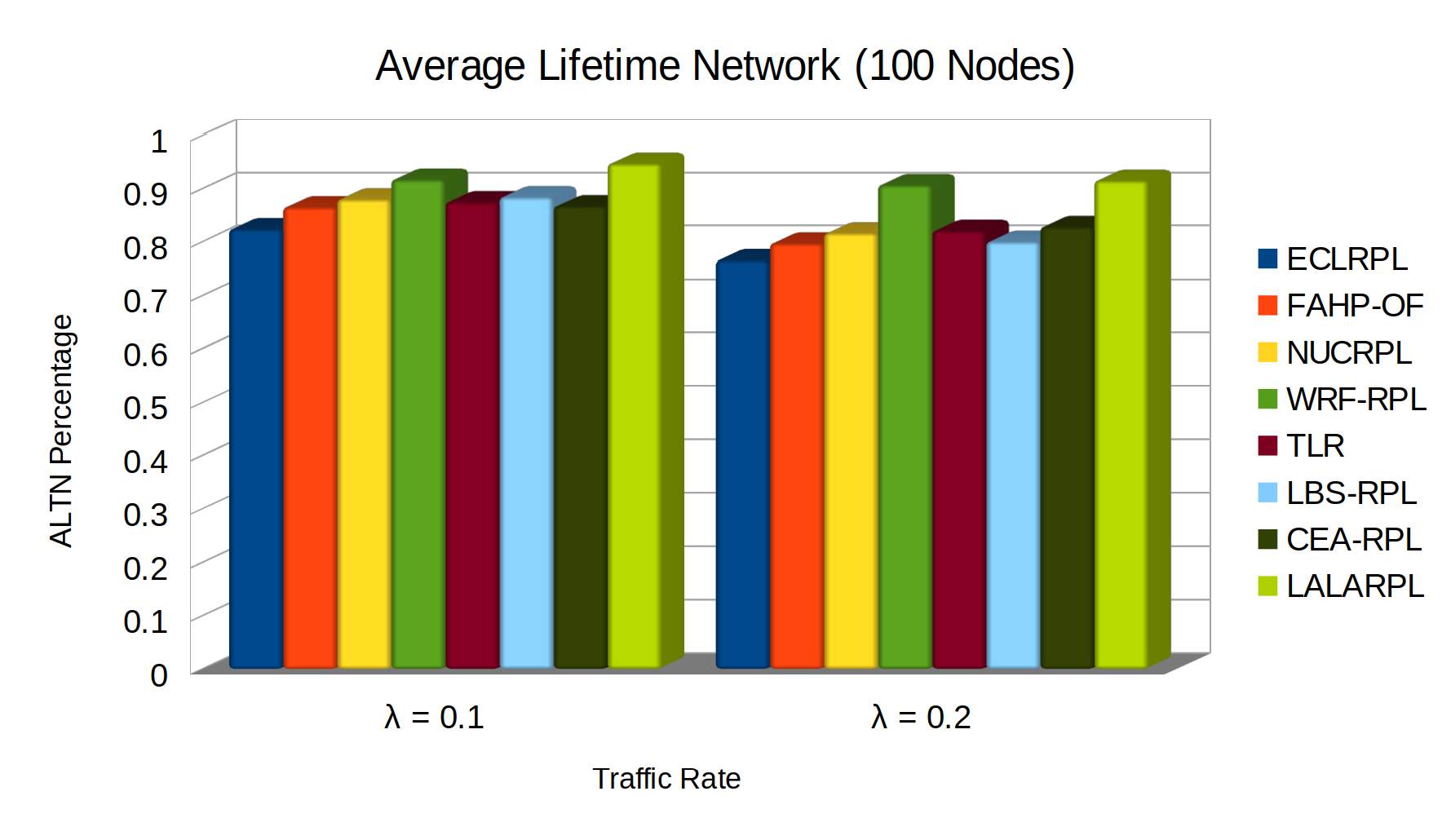}
    \caption{Average Lifetime Network in 100 nodes test}
    \label{fig:Average-Lifetime-Network-100}
\end{figure}

\begin{figure}
    \centering
    \includegraphics[width=0.5\linewidth]{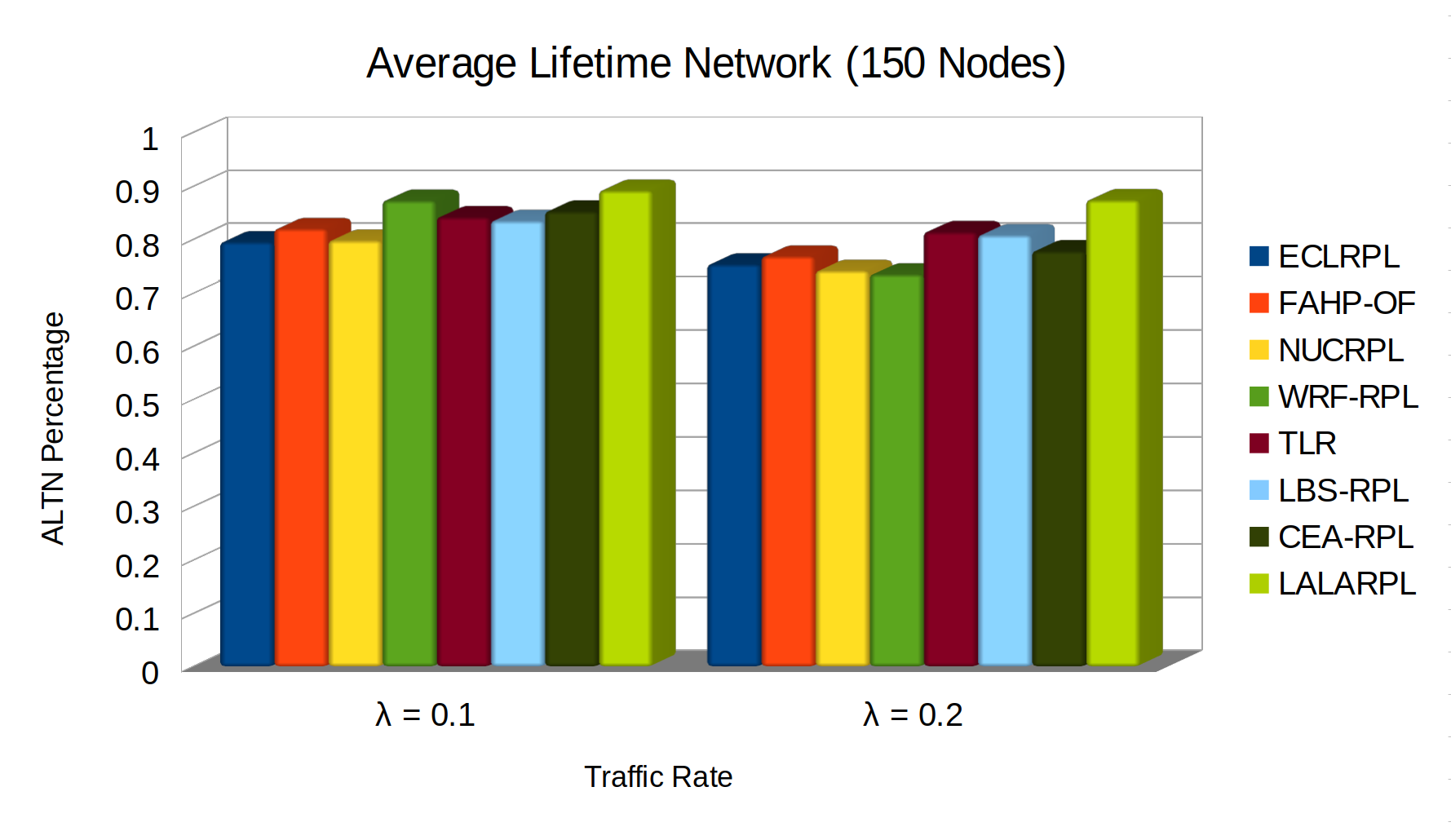}
    \caption{Average Lifetime Network in 150 nodes test}
    \label{fig:Average-Lifetime-Network-150}
\end{figure}

Analyzing the simulation results for the LALARPL method reveals that the protocol excels at extending the operational lifetime of IoT devices. Figures \ref{fig:Average-Lifetime-Network-50}, \ref{fig:Average-Lifetime-Network-100}, and \ref{fig:Average-Lifetime-Network-150} correspond to the simulation results of networks with 50, 100, and 150 nodes, respectively, and visually represent the ALTN across these scenarios.

The LALARPL method's performance can be attributed to several key factors:

\begin{itemize}
    \item Efficient Energy Management: By intelligently limiting the communication and processing tasks per node, LALARPL prevents premature energy depletion, which directly translates to a longer ALTN.
    \item Learning Automata: The protocol's use of learning automata ensures that routing decisions are optimized in real-time. This adaptability prevents nodes from expending unnecessary energy, especially in high-traffic conditions, thereby delaying the time to the first node's death.
    \item Load Balancing: LALARPL's load balancing mechanisms evenly distribute the energy demands across the network, ensuring no single node bears excessive burden. This prevents the formation of energy hotspots and results in a longer average node lifetime.
\end{itemize}

For the 50-node network at $\lambda = 0.1$ packets/s, LALARPL shows a superior ALTN of 0.975, compared to 0.958 by the following best protocol, WRF-RPL, highlighting a $1.78\%$ improvement. When the packet interval increases to $\lambda = 0.2$ packets/s, LALARPL achieves a $2.52\%$ increase in ALTN compared to WRF-RPL's 0.921. With the 100-node network at $\lambda = 0.1$ packets/s, LALARPL's ALTN stands at 0.948, significantly higher than WRF-RPL's 0.918, marking a $3.27\%$ improvement. At the higher packet interval of $\lambda = 0.2$ packets/s, the ALTN improvement is evident at $0.88\%$ over WRF-RPL's 0.908. For the most extensive network tested, consisting of 150 nodes, at $\lambda = 0.1$ packets/s, LALARPL registers an ALTN of 0.893, which is $2.17\%$ better than WRF-RPL's 0.874. At $\lambda = 0.2$ packets/s, LALARPL maintains its lead with an ALTN of 0.875, an impressive $19.04\%$ increase over WRF-RPL's 0.735.

These outcomes underscore the effectiveness of the LALARPL method in enhancing the longevity of the network by ensuring that the first node and subsequent nodes have a significantly delayed time of death. The delay in energy depletion of the first node is particularly crucial, as it indicates a resilient and well-distributed network that can sustain its operational capabilities for extended periods, which is critical to IoT applications. The enhanced ALTN achieved by LALARPL affirms the protocol's advanced energy management capabilities, showcasing its potential to improve the sustainability of IoT networks. This is particularly beneficial in scenarios where network reliability and extended operation without maintenance are of the essence.

\clearpage

\clearpage

\section{Conclusion}\label{sec6}
This paper presented an enhanced load-balancing algorithm for the Routing Protocol for LLNs (RPL) that employs learning automata, which is designated LALARPL. Our innovative approach significantly optimizes routing in Internet of Things (IoT) environments by integrating advanced decision-making capabilities that dynamically adapt to network conditions. The key benefits of LALARPL, as demonstrated through extensive simulation, include improved packet delivery ratios, reduced end-to-end delays, and more efficient energy usage across the network. By adopting a traffic-aware and energy-efficient strategy, LALARPL ensures a fair distribution of network load, extending the lifetime of individual nodes and enhancing the overall performance and reliability of IoT networks. Moreover, the learning automata component allows for the adaptive adjustment of routing decisions based on real-time network feedback, which optimizes both the throughput and the energy consumption. This feature is crucial in maintaining IoT devices' sustainability and operational efficiency, often limited by battery life and processing power. In conclusion, the introduction of LALARPL represents a significant step forward in the evolution of RPL-based networking solutions, providing a robust framework for achieving optimal load balancing and efficient resource management in IoT networks. The results suggest that LALARPL can serve as a foundational technology for future developments in IoT routing protocols, potentially leading to more resilient and efficient network architectures tailored to the unique demands of the burgeoning IoT landscape.

As part of future work, introducing mobility management into the LALARPL framework presents a promising avenue for research. Mobility in IoT environments introduces additional challenges, including varying network topology, increased packet loss, and dynamic routing paths that require real-time adaptation. Integrating mobility would allow LALARPL to dynamically adjust to changing network conditions as nodes move, maintaining efficient routing and load balancing. Future iterations could incorporate mobility models that simulate real-world IoT applications, such as vehicular networks or mobile health devices, to assess the performance of LALARPL in managing dynamic and heterogeneous networks. Moreover, developing algorithms that predict node movements and preemptively adjust routing decisions could further enhance the robustness and efficiency of the RPL protocol in mobile scenarios. This mobility-aware version of LALARPL would be instrumental in broadening the applicability of RPL to a broader array of IoT systems where node mobility is a key characteristic. This exploration would contribute to the theoretical advancements in routing protocols and have practical implications for deploying IoT solutions in urban settings, disaster response scenarios, and other dynamic environments where mobility plays a crucial role.

\clearpage

\section{References}\label{sec7}


\newpage
\section*{Author Biographies}\label{sec8}

\noindent\begin{tabular}{m{0.25\textwidth} m{0.7\textwidth}}
    \adjustbox{valign=t}{\includegraphics[width=0.25\textwidth]{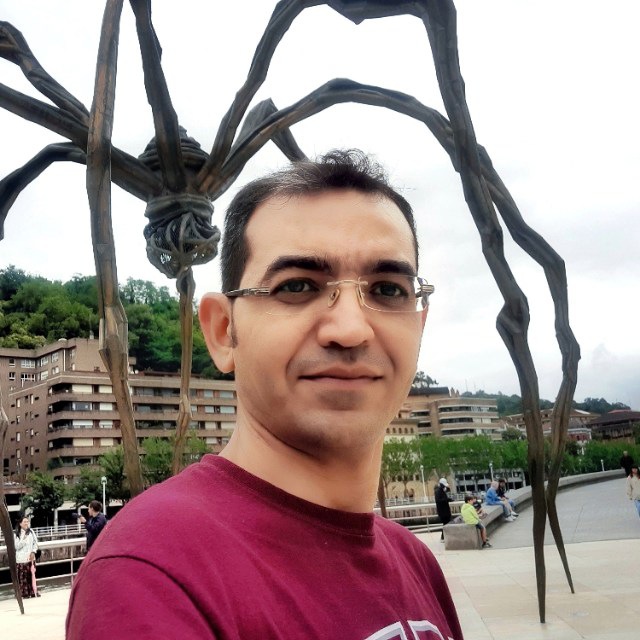}} &
    \textbf{MOHAMMADHOSSEIN HOMAEI (M'19)} was born in Hamedan, Iran. He obtained his B.Sc. in Information Technology (Networking) from the University of Applied Science and Technology, Hamedan, Iran, in 2014 and his M.Sc. from Islamic Azad University, Malayer, Iran, in 2017. He is pursuing his Ph.D. at Universidad de Extremadura, Spain, where his prolific research has amassed over 100 citations.
    
    Since December 2019, Mr. Homaei has been affiliated with Óbuda University, Hungary, as a Visiting Researcher delving into the Internet of Things and Big Data. His tenure at Óbuda University seamlessly extended into a research collaboration with J. Selye University, Slovakia, focusing on Cybersecurity from January 2020. His research voyage led him to the National Yunlin University of Science and Technology, Taiwan, where he was a Scientific Researcher exploring IoT and Open-AI from January to September 2021. His latest role was at the Universidade da Beira Interior, Portugal, in the Assisted Living Computing and Telecommunications Laboratory (ALLab), from June 2023 to January 2024, where he engaged in cutting-edge projects on digital twins and machine learning. He is the author of ten scholarly articles and holds three patents, highlighting his diverse research interests in Digital Twins, Cybersecurity, Wireless Communications, and IoT. 
    
    An active IEEE member, Mr. Homaei has carved a niche for himself with notable contributions to Digitalization, the Industrial Internet of Things (IIoT), Information Security Management, and Environmental Monitoring. His substantial work continues to influence the technological and cybersecurity landscape profoundly. 
\end{tabular}

\newpage

\appendix
\section{Appendix A: Lemma and Proof of Convergence in Learning Automata}

\textbf{Lemma:} 
\textit{Consider a set of parent nodes \(PS = \{P_1, P_2, \dots, P_N\}\) for a given child node, where each parent node \(P_i\) has an associated traffic index \(TI_i\) and selection probability \(P_i\) determined by a Learning Automaton. Under the reward and penalty scheme defined in the algorithm, the selection probability \(P_{i^*}\) for the optimal parent node \(P_{i^*}\) converges to 1, where \(P_{i^*}\) minimizes the overall network load and balances traffic distribution.}

\textbf{Proof:}

1. \textit{Initial Setup:} \\
Consider the set of parent nodes \(PS = \{P_1, P_2, \dots, P_N\}\) for a child node. The selection probability \(P_i\) for each parent node \(P_i\) is given by:

\[
P_i(t+1) = P_i(t) + \alpha(t) \left[ r_i(t) - P_i(t) \right]
\]

where:
\begin{itemize}
    \item \(P_i(t)\) is the probability of selecting parent node \(P_i\) at iteration \(t\),
    \item \(\alpha(t)\) is the learning rate at iteration \(t\),
    \item \(r_i(t)\) is the reward indicator for parent node \(P_i\) at iteration \(t\), with \(r_i(t) = 1\) if a reward is received and \(r_i(t) = 0\) if a penalty is received.
\end{itemize}

2. \textit{Learning Automata Update Rule:} \\
The Learning Automata updates the selection probability \(P_i(t)\) based on the received feedback (reward or penalty). The reward is assigned according to the following rules:
\begin{itemize}
    \item \(r_i(t) = 1\) (reward) if \(TI_i(t) < 0.5 \cdot \text{avg}(TI_j(t))\) for \(j \in PS\),
    \item \(r_i(t) = 1\) (reward) if \(0.5 \cdot \text{avg}(TI_j(t)) \leq TI_i(t) < 0.8 \cdot \text{avg}(TI_j(t))\) and \(\text{numhop}_i = \min\{\text{numhop}_j\}\) for \(j \in PS\),
    \item \(r_i(t) = 0\) (penalty) if \(TI_i(t) > \text{avg}(TI_j(t))\) for \(j \in PS\).
\end{itemize}

3. \textit{Convergence Analysis:} \\
The selection probability \(P_i(t)\) evolves over time according to the following recursive relation:

\[
P_i(t+1) = P_i(t) + \alpha(t) \left[ r_i(t) - P_i(t) \right]
\]

This update rule is typical in reinforcement learning, where the learning rate \(\alpha(t)\) decreases over time to ensure convergence. The algorithm is designed so that the selection probability \(P_i(t)\) will increase for nodes with favorable traffic indices (i.e., those receiving rewards) and decrease for nodes with high traffic indices (i.e., those receiving penalties).

As \(t \to \infty\), the Learning Automata will converge to an optimal selection strategy where the probability \(P_{i^*}(t) \to 1\) for the parent node \(P_{i^*}\) that consistently receives rewards, indicating it has the best combination of low traffic index and proximity to the network root.

4. \textit{Optimality Criterion:} \\
The parent node \(P_{i^*}\) that achieves the maximum selection probability minimizes the overall network load by maintaining a balanced traffic distribution. The conditions for the reward ensure that \(P_{i^*}\) will have the lowest traffic index relative to other parent nodes while also being close to the network root, ensuring efficient data transmission.

Thus, the lemma is proved, demonstrating that the Learning Automata converges to the optimal parent node, \(P_{i^*}\), ensuring efficient and balanced network performance.

\end{document}